\documentclass[prb,aps,twocolumn,superscriptaddress,floatfix,showpacs]{revtex4-2}
  \usepackage{environ}
\usepackage[colorlinks=true, citecolor=blue, linkcolor=blue, urlcolor=blue]{hyperref}

\usepackage{graphicx} % Required for inserting images
\usepackage{amssymb}
\usepackage{amsmath}
\usepackage{bbold}
\usepackage{mathtools}
\usepackage[dvipsnames]{xcolor}
\usepackage{float}					% Easy positioning of figures
\usepackage{ dsfont }   			    % For double stroke letters, like for real or complex number sets
\usepackage{notes2bib}				% Send notes to Reference
\usepackage{comment}
\usepackage[normalem]{ulem} 
\usepackage{braket}

\bibnotesetup{
	note-name = ,
	use-sort-key = false
}

\graphicspath{{./plots/}}

\begin{document}
\title{Local Information Flow in Quantum Quench Dynamics}

\author{Nicolas P. Bauer}
\email[Email:\,]{nicolas.bauer@uni-wuerzburg.de}
\affiliation{ Institute of Theoretical Physics and Astrophysics, University of Würzburg, Germany} 
\affiliation{Würzburg-Dresden Cluster of Excellence ct.qmat, Germany}
\author{Björn Trauzettel}
\affiliation{ Institute of Theoretical Physics and Astrophysics, University of Würzburg, Germany} 
\affiliation{Würzburg-Dresden Cluster of Excellence ct.qmat, Germany}
\author{Thomas Klein Kvorning}
\affiliation{ Department of Physics, KTH Royal Institute of Technology, Stockholm, 106 91 Sweden} 
\author{Jens H. Bardarson}
\affiliation{ Department of Physics, KTH Royal Institute of Technology, Stockholm, 106 91 Sweden} 
\author{Claudia Artiaco}
\email[Email:\,]{artiaco@kth.se}
\affiliation{ Department of Physics, KTH Royal Institute of Technology, Stockholm, 106 91 Sweden}

\begin{abstract}
    We investigate the out-of-equilibrium dynamics of quantum information in one-dimensional systems undergoing a quantum quench using a local perspective based on the information lattice.
    This framework provides a scale- and space-resolved decomposition of quantum correlations, enabling a hydrodynamic description of the information flow through well-defined local densities---termed local information---and currents.
    We apply this framework to three local quenches in noninteracting fermionic chains: (i)~the release of a single particle into an empty tight-binding chain, (ii)~the connection of two critical chains via the removal of a central barrier, and (iii)~the coupling of a topological Kitaev chain to a critical chain.
    In each case, the information lattice reveals the local structure of correlation buildup and information interface effects, going beyond global measures such as the von Neumann entropy.
    In particular, through the information lattice we uncover the signatures in the local information flow associated with topological edge modes and analytically explain the fractional von Neumann entropy values observed in Majorana quench protocols.
    Our approach is general and applicable to interacting, disordered, and open systems, providing a powerful tool for characterizing quantum information dynamics.
\end{abstract}

\maketitle

\section{Introduction}

Recent experimental breakthroughs in atomic physics, quantum optics, and nanoscience have enabled the realization and control of highly tunable quantum many-body systems, establishing out-of-equilibrium dynamics as a central topic in both theoretical and experimental condensed matter research~\cite{polkovnikov2011colloquium,eisert2015quantum,defenu2024out}.
The synergy between theory and experiment has shed new light on fundamental questions---such as how thermalization occurs in closed quantum systems~\cite{neumann2010proof}---and is key to developing quantum technologies, including quantum computing and quantum simulation, where precise real-time control of quantum states is crucial.
One of the most studied protocols for inducing out-of-equilibrium dynamics is the quantum quench, where a system is prepared in an eigenstate $|\psi_0\rangle$ of a Hamiltonian $H_0$ and subsequently evolved under a different Hamiltonian $H$ for which $|\psi_0\rangle$ is not an eigenstate~\cite{farhi2000quantum,lauchli2008spreading,fagotti2008evolution,rigol2009quantum,altshuler2010anderson,chen2011quantum,calabrese2012quantum,trotzky2012probing,gamayun2020fredholm,gnezdilov2023ultrafast,ohanesjan2023energy}.

Universal features of the post-quench dynamics are captured by quantum information quantities, such as the scaling of the von Neumann entropy~\cite{calabrese2004entanglement,calabrese2005evolution,dechiara2006entanglement,calabrese2007quantum,kim2013ballistic} and other entanglement measures \cite{coser2014entanglement,murciano2022quench,jafari2025entanglement} with time.
Quantum information tools abstract away system-specific physical properties.
In particular, the dynamics of the von Neumann entropy $S_A$ of a region $A$ reveals how correlations spread between $A$ and the rest of the system.
In generic clean one-dimensional systems with local interactions and with approximately homogeneous initial states, $S_A$ increases linearly with time before saturating to a value proportional to the size of $A$~\cite{calabrese2005evolution,dechiara2006entanglement,calabrese2007quantum,kim2013ballistic}.
The stationary von Neumann entropy in the infinite time limit has the same density as the thermodynamic entropy, implying that local multi-point correlation functions may be evaluated as averages over the thermal ensemble or its generalizations~\cite{rigol2007relaxation,alba2017entanglement}.
Through the holographic principle and the Ryu-Takayanagi formula~\cite{ryu2006holographic}, the study of quantum information dynamics has a broad relevance also in the context of general relativity and black holes~\cite{nozaki2013holographic}.

Given the difficulty in carrying out ab initio calculations of quantum information dynamics, simple heuristic pictures have been developed.
Examples are the entanglement quasiparticle picture~\cite{calabrese2006time,calabrese2016introduction,calabrese2020notes}, where correlations are spread by pairs of entangled quasiparticles with opposite momenta, the membrane picture~\cite{nahum2017quantum,jonay2018coarse,zhou2020entanglement}, and the entanglement tsunami~\cite{liu2014entanglement}.
These pictures are based on the description of the quantum information dynamics at the local level, which provides deeper insights than global quantities such as the von Neumann entropy.
Other approaches in this direction include the entanglement link representation~\cite{singha2020entanglement,singha2021link,santalla2023entanglement} and the entanglement contour~\cite{chen2014entanglement,kudlerflam2019holographic}.

This local perspective can in principle reveal novel fundamental aspects of information dynamics---for example, information interface effects that might arise in setups relevant to quantum transport where two systems with different quantum information properties, such as critical and noncritical systems, are suddenly connected~\cite{eisler2007evolution,calabrese2007entanglement,eisler2008entanglement,stephan2011local,eisler2014area}, as well as the peculiar features of quantum information transport in topological systems~\cite{smirnov2015majorana,sela2019detecting,bauer2023quench}.
However, constructing a universal framework that is capable of decomposing quantum information at the local level systematically at any evolution time is nontrivial since quantum information is intrinsically nonlocal.
Recent works~\cite{artiaco2024universal,klein2022time,artiaco2024efficient,harkins2025nanoscale} demonstrated that the \textit{information lattice} provides precisely such a decomposition for one-dimensional systems with open boundary conditions.
The information lattice defines \textit{local information}, which quantifies the total amount of correlations in a region on a given scale and spatial location that cannot be found in any smaller subregions.
Through the information lattice, the total information in the system becomes a hydrodynamic quantity characterized by well-defined local densities (local information) and currents.
The out-of-equilibrium dynamics of information is cast into the flow of local information within nearby information lattice sites, akin to the flow of a local charge.
The information lattice has recently been employed to both define a universal framework for classifying quantum states and their characteristic length scales based on their scale-resolved correlation structure, as demonstrated for ground and midspectrum eigenstates of the disordered interacting Kitaev chain~\cite{artiaco2024universal}, and to obtain accurate approximate dynamics of local observables in large-scale interacting quantum systems~\cite{klein2022time,artiaco2024efficient,harkins2025nanoscale}.

In this article, we demonstrate the use of the local information flow within the information lattice to fully uncover the dynamics of quantum information in a quantum quench.
We focus on three local quench protocols in noninteracting fermionic chains organized in an increasing order of complexity to showcase the characterization of quench dynamics through the local information flow.
After providing the definition and describing the properties of the information lattice in Sec.~\ref{sec:information-lattice}, we introduce our framework in Sec.~\ref{sec:potential-well-quench} by investigating a simple local quench in which a single particle is released into an empty tight-binding chain.
We follow the dynamics of the particle in terms of the propagation of both short- and long-range local information packets, giving us a fully local interpretation of the information dynamics, and connect these to the real-space picture.
In Sec.~\ref{sec:critical-states-quench}, we consider a single-site quench in a critical chain.
In critical systems, information is present at all scales; thus, they act as ``reservoirs'' of local information~\cite{beenakker2004quantum}.
We characterize novel information interface phenomena in this quench protocol.
In Sec.~\ref{sec:majorana-teleportation-quench}, we analyze the flow of local information when a topological Kitaev chain is suddenly connected to a critical tight-binding chain and obtain a comprehensive description of the physical mechanisms governing the dynamics through the information lattice.
The Kitaev chain hosts a pair of Majorana edge modes forming a delocalized fermionic mode, which can be occupied at zero energy cost~\cite{kitaev2001unpaired,lathinen2017ashortintroduction,alicea2012new}.
While topologically protected, the delocalized fermionic mode is susceptible to decoherence---even under local coupling to just one of the constituent Majoranas~\cite{goldstein2011decay,budich2012failure}.
We use the information lattice to systematically characterize this decoherence process.
Additionally, the information lattice allows us to derive simple analytical arguments that fully explain the fractional von Neumann entropy signatures previously observed in similar quench setups~\cite{sela2019detecting,bauer2023quench}.

The quenches examined in this article serve as illustrative examples focusing on local quench protocols within noninteracting (topological) fermionic systems.
However, our framework is universal and applicable to any system, including interacting, disordered, or dissipative chains.
In Sec.~\ref{sec:conclusions}, we summarize our findings and discuss potential future applications.

\begin{figure}[t]
	\centering
	\includegraphics[width=\columnwidth]{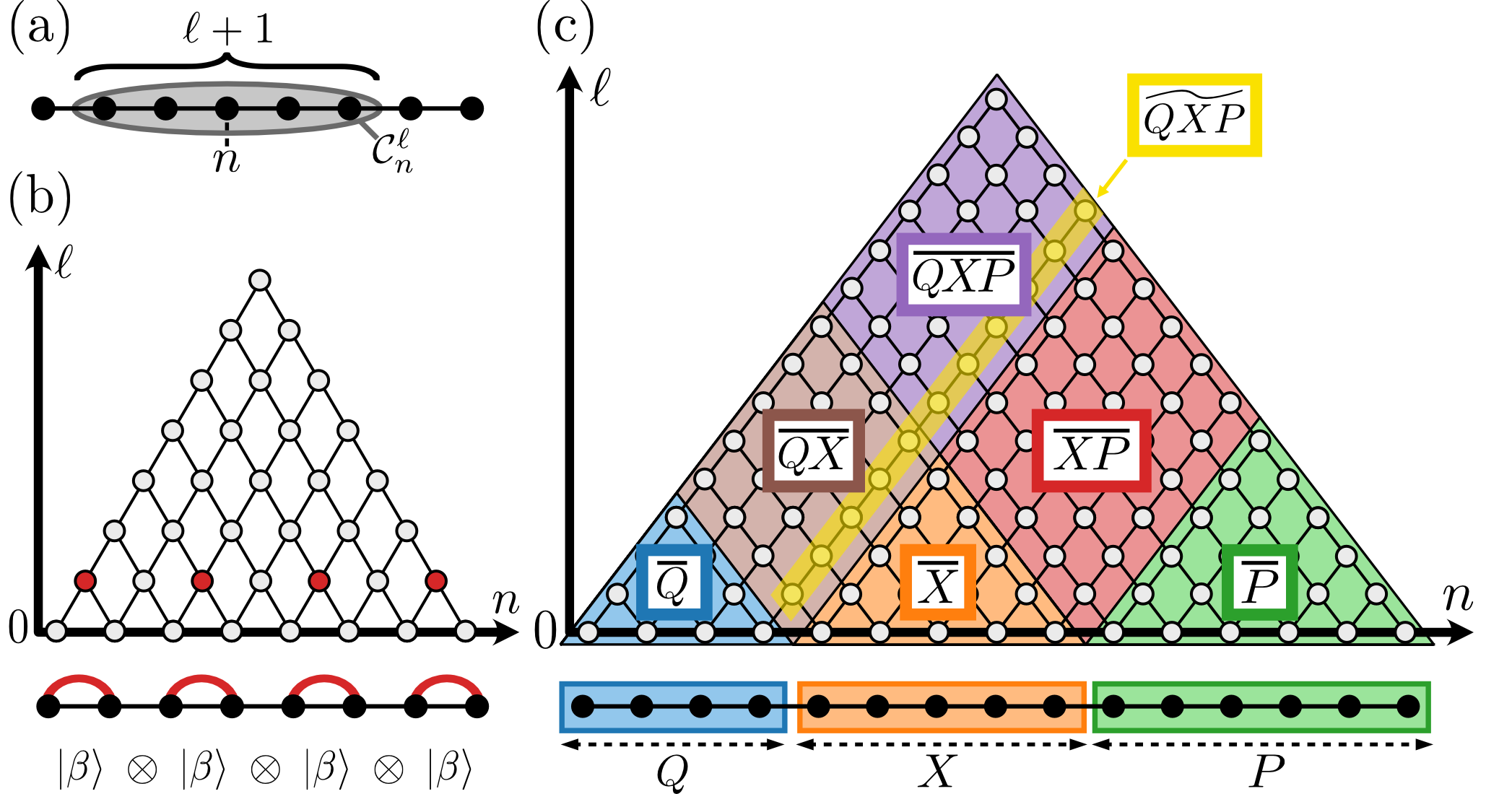}
	\caption{(a) Illustration of the subsystem $\mathcal{C}^\ell_n$ encompassing $\ell + 1$ physical sites centered around $n$.
    (b) Schematic of the information lattice.
    Shown is the information lattice for a Bell-pair product state on nearest-neighbor sites (bottom black-filled circles connected by red lines) for $N=8$.
    Colors quantify local information: bright red indicates $i^\ell_n=2$; gray dots correspond to $i^\ell_n=0$.
    Note that for an even (odd) scale $\ell$, the lattice sites $n$ takes on integer (half-integer) values.
    Nonzero local information is present exclusively at $\ell=1$ on every second lattice site $n=\{\frac{1}{2},\frac{5}{2},\frac{9}{2},\frac{13}{2}\}$, which corresponds to the subsystem Bell pairs.
    (c) Schematic of an information lattice partition for a physical chain divided into three nonoverlapping regions: $Q$, $X$, and $P$.
    The background colors represent the information lattice partition corresponding to these regions.
    The interface $\widetilde{QXP}$ between regions $Q$ and $XP$ is highlighted in yellow.
    Gray circles indicate the information lattice sites.
    For demonstration purposes, we illustrate the information lattice for a fully mixed state $\rho \propto \mathbb{1}$ in which $i^\ell_n = 0$ for all $(n, \ell)$.
    }
	\label{fig:IL-Partition}
\end{figure}

\section{Information lattice}
\label{sec:information-lattice}

\subsection{Definition}

In this section, we define the information lattice, which we then use to analyze the time evolution of local information in various quantum-quench protocols.
For simplicity, we assume that the physical sites have only two quantum degrees of freedom (qubits).
The von Neumann information (or total information) $I(\rho)$ in a quantum state $\rho$ equals the deficit of the von Neumann entropy $S(\rho)$ from its maximum:
\begin{align}
    I(\rho) &= \log_2[\dim(\rho)] - S(\rho) \notag \\
    &= N + \mathrm{Tr}[\rho \log_2(\rho)],
    \label{eq:von_neumann_information}
\end{align}
where $\dim(\rho)$ is the Hilbert space dimension of the entire system and $N$ is the total number of qubits.
$I(\rho)$ is the information stored in the state $\rho$ concerning measurement outcomes, that is, it is the additional predictive power gained by knowing the state $\rho$ compared to assuming a maximally mixed state~\footnote{
Specifically, $I(\rho)$ is the additional average number of binary outcomes per measurement that can be predicted with certainty in the asymptotic limit of infinitely many measurements~\cite{shannon1948bell}.
For example, if exactly one binary-outcome measurement can be predicted with certainty---as in the case of a qubit in a pure state---then $I(\rho)=1$ bit.}.
Analogously, the von Neumann information in the reduced density matrix $\rho_A = \mathrm{Tr}_{\bar{A}}(\rho)$ of the subsystem $A$,
\begin{equation}
    I(\rho_A) = N + \mathrm{Tr}[\rho_A \log_2(\rho_A)],
\end{equation}
is the information stored in $\rho_A$.

We define local information as the decomposition of the total information in the state $I(\rho)$ into local contributions at each scale and spatial location.
In a one-dimensional system under open boundary conditions, such a decomposition is constructed in the following way.
We decompose the chain in all possible subsystems $\mathcal{C}^\ell_n$ made of $\ell+1$ neighboring sites centered around position $n$ such that subsystems with $\ell=0$ are the physical sites, as illustrated in Fig.~\ref{fig:IL-Partition}(a).
The reduced density matrix of subsystem $\mathcal{C}^\ell_n$ is $\rho^\ell_n = \mathrm{Tr}_{\bar{\mathcal{C}}_n^{\ell}}(\rho)$.
The local information $i^\ell_n$ is then defined as the decomposition of the total information in $\rho^\ell_n $ for any $(n,\ell)$,
\begin{equation}
\label{eq:def-local-information-decomposition}
    I(\rho^\ell_n) = \sum_{(n^\prime,\ell^\prime) \in \mathcal{D}^\ell_n} i^{\ell^\prime}_{n^\prime},
\end{equation}
with $\mathcal{D}^\ell_n = \{ (n^\prime, \ell^\prime) | \mathcal{C}^{\ell^\prime}_{n^\prime} \subseteq \mathcal{C}^\ell_n \}$.
This gives
\begin{eqnarray}
\label{eq:def-local-information-mutual-information}
    i^\ell_n = I(\rho^\ell_n) - I(\rho^{\ell-1}_{n-1/2}) - I(\rho^{\ell-1}_{n+1/2}) + I(\rho^{\ell-2}_{n}),
\end{eqnarray}
where it is implicit that the von Neumann information of empty subsystems is zero.
The local information $i^\ell_n$ quantifies how much more we can predict about measurement outcomes by knowing the density matrix $\rho^\ell_n$ rather than the density matrices $\rho^{\ell-1}_{n-1/2}$ and $\rho^{\ell-1}_{n+1/2}$ of the smaller-scale subsystems contained in $\mathcal{C}^\ell_n$; it follows that $i^\ell_n \geq 0$.
As an example, consider a two-site system in a Bell pair state: $ \lvert \beta \rangle = 1/\sqrt{2} (\lvert \uparrow \downarrow \rangle + \lvert \downarrow \uparrow \rangle )$.
The single-site density matrices are $\rho^0_0 = \rho^0_1 = \frac{1}{2} \mathbb{1}$; thus, $i^0_0 = i^0_1 = 0$.
This formalizes that knowing the system is in a Bell pair state offers no predictive power for single-site measurements: each of the two possible outcomes occurs with equal probability $1/2$.
In contrast, $i^1_{1 / 2} = 2$ shows that the total information in the state is accessible by measurements of two-site operators.

The decomposition~\eqref{eq:def-local-information-decomposition} defines the information lattice---a hierarchical triangular structure in which sites are labeled by indexes $(n,\ell)$ and are associated with local information $i^\ell_n$ in Eq.~\eqref{eq:def-local-information-mutual-information}.
Fig.~\ref{fig:IL-Partition}(b) illustrates the information lattice for an example state given by the product of Bell pairs $\lvert \beta \rangle$ between nearest-neighbor sites.
Gray dots contain zero local information while red dots correspond to 2 bits.
As $i^\ell_n$ precisely accounts for the information present only at a given scale and spatial location, in this example nonzero local information is present only at $\ell=1$ on every second lattice site.
Information lattice sites with $\ell > 1$ are associated with zero local information, as there are no correlations shared between sites belonging to different Bell pairs.

Notice that the information lattice is also well-defined for mixed states.
The only distinction is that the total information in the system is less than that of a pure state, that is, $I(\rho) < \log_2 \left[ \dim(\rho) \right]$.
However, the definition of local information~\eqref{eq:def-local-information-mutual-information} and all its associated properties (see Sec.~\ref{sec:info-lattice-properties}) remain unchanged.

\subsection{Properties}
\label{sec:info-lattice-properties}

The total information in the system $I(\rho)$ defined in Eq.~\eqref{eq:von_neumann_information} is conserved under unitary time evolution.
As a result, definition~\eqref{eq:def-local-information-decomposition} implies that the sum of local information within the whole information lattice is constant: $I(\rho) = \sum_{(n, \ell) \in \mathcal{D}} i^{\ell}_{n} = \mathrm{const.}$, with $\mathcal{D} = \{ (n, \ell) \, | \, \mathcal{C}^{\ell}_{n} \subseteq \mathcal{S} \}$ and $\mathcal{S}$ the entire physical chain.
The decomposition of the total information performed by the information lattice makes $I(\rho)$ akin to a hydrodynamic conserved quantity with well-defined local densities (that is, $i^\ell_n$) and local currents.
Local information currents are defined similarly to those of a locally conserved operator in a system with a local Hamiltonian.
Their explicit form can be derived from the von Neumann equation of motion for subsystem density matrices~\cite{klein2022time,artiaco2024efficient}.
Under the unitary time evolution governed by a local Hamiltonian, these currents propagate through the information lattice along the diagonal black lines shown in Fig.~\ref{fig:IL-Partition}(b)-(c).
In conclusion, the information lattice acts as an ``information microscope'', revealing the fine-grained structure of correlations and their time evolution across different scales and spatial locations.

The local perspective provided by the information lattice extends conventional quantum information approaches that study information transport using global quantities, such as bipartite von Neumann entropy and mutual information~\cite{nielsen2010quantum,wilde2013quantum}.
Global approaches are limited in several ways compared to the information lattice.
For instance, by measuring the von Neumann entropy $S(\rho_A)$ of a region $A$, one cannot know with which part of the complement region $A$ is entangled.
In other words, $S(\rho_A)$ does not provide any knowledge about correlations at scales that exceed the size of the region $A$, and the length of $A$ defines the maximum scale of information that can be detected.
As a second example, the mutual information cannot directly be used to quantify at which scale correlations are; the mutual information between two disjoint regions $A$ and $B$ exactly captures the correlations at the scale corresponding to the distance between $A$ and $B$ only if the state of the entire system is in a product state between $A \cup B$ and the rest of the system.
In general, estimating the information at a certain scale using the mutual information between disjoint regions can lead to either an underestimation or an overestimation of the correlations~\cite{artiaco2024universal}.

\subsection{Information lattice partitions according to regions $Q$, $X$, and $P$}

By summing local information within two-dimensional partitions of the information lattice, well-defined global information quantities can be extracted.
This is again a consequence of definition~\eqref{eq:def-local-information-decomposition}, which implies that different nonoverlapping partitions of the information lattice contain information of independent scale and spatial regions of the system.
Following Ref.~\cite{bauer2023quench}, for the quench protocols discussed in this article we consider the physical chain as divided into three regions: $Q$ composed of $l_Q$ physical sites; $X$ made of $l_X$ physical sites on the right of $Q$; and $P$ of length $l_P$ on the right of $X$.
$Q$ contains the quenched physical site(s).
The separation layer $X$ provides a knob to investigate the time of flight, tails, and interface effects in the local information flow.
$P$ denotes a probe region to test the development of correlations within the system over time.
Given $Q$, $X$, and $P$, the information lattice is naturally decomposed according to the partition illustrated in Fig.~\ref{fig:IL-Partition}(c) by different background colors.
The equilateral triangles $\overline{Q}$, $\overline{X}$, and $\overline{P}$ encompass only information that is completely accessible by knowing the reduced density matrices of $Q$, $X$, and $P$.
On the other hand, $\overline{QX}$, $\overline{XP}$, and $\overline{QXP}$ contain information that is accessible only by knowing the states of the enlarged regions $QX$, $XP$, and $QXP$.
By indicating the information lattice partitions as $\Lambda \in  \{ \overline{Q},\overline{X},\overline{P},\overline{QX},\overline{X P},\overline{QXP} \}$, we define the change of total information in $\Lambda$ as 
\begin{equation}
        \Gamma_\Lambda(t) = \sum_{(n, \ell) \in \Lambda} \left[ i_n^\ell(t) - i_n^\ell(0^-) \right],
    \label{eq:totalInformation}
\end{equation}
where the sum runs over all information lattice sites $(n,\ell)$ within the partition $\Lambda$, and $i^\ell_n(0^-)$ is the local information before the quench is performed.

From Eq.~\eqref{eq:def-local-information-decomposition}, summing local information $i^\ell_n$ within any of the information lattice partitions $\overline{Q}$, $\overline{X} $ and $ \overline{P}$ gives the total information in the subsystems $Q$, $X$ and $P$, respectively.
Thus, $\Gamma_\Lambda$ for $\Lambda = \{ \overline{Q}, \overline{X}, \overline{P} \}$ is the difference in the total information in $Q$, $X$ and $P$ with respect to the total information stored in their reduced density matrices in the initial state.
Summing local information $i^\ell_n$ in the larger-scale partition $\overline{QX}$ yields how much more one can predict about measurement outcomes by knowing the reduced density matrix $\rho_{QX}$ of subsystem $QX$ compared to knowing only $\rho_Q$ and $\rho_X$ of subsystems $Q$ and $X$.
In other words, $\sum_{(n,\ell) \in \overline{QX}} \, i^\ell_n$ is the mutual information between the subsystems $Q$ and $X$.
Similarly for $\Gamma_{\overline{XP}}$.
Finally, $\Gamma_{\overline{QXP}}$ is the additional information we have access to from the full system density matrix $\rho_{QXP}$ compared to the density matrices $\rho_{QX}$ and $\rho_{XP}$ of the overlapping subsystems $QX$ and $XP$.
Fig.~\ref{fig:IL-Partition}(c) also shows the $\widetilde{QXP}$ interface between regions $Q$ and $XP$ highlighted in yellow.
$\widetilde{QXP}$ contains the correlations between the rightmost physical site of $Q$ and the $X$ and $P$ regions.

\section{Releasing one particle in an empty tight-binding chain}
\label{sec:potential-well-quench}

\begin{figure}[tbp]
	\centering
	\includegraphics[width=\columnwidth]{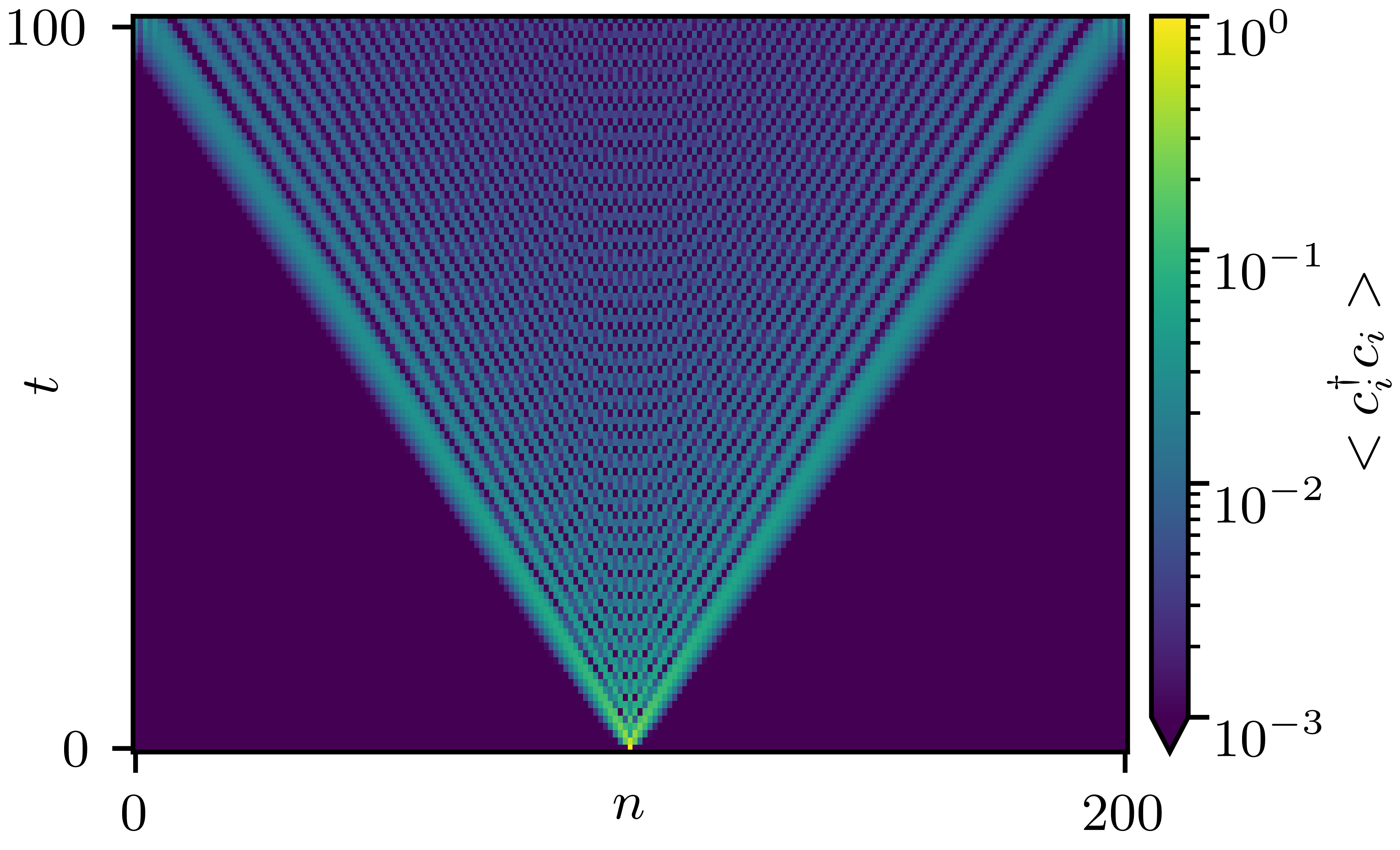}
	\caption{Time evolution of the occupation density after releasing a particle at the central site $\eta=100$ in an empty tight-binding chain with $N=201$ sites.
    Before the quench, the system is in the ground state of the Hamiltonian~\eqref{eq:HamTightBinding} with $\mu_i = -20 \tau_p$ and $\mu_p = 20 \tau_p$.
    After the quench, the dynamics is governed by the Hamiltonian~\eqref{eq:HamTightBinding} with $\mu_f = \mu_p = 20 \tau_p$.
    }
	\label{fig:occupation_density_NonCrit_TimeEvo}
\end{figure}

\begin{figure*}[tbp]
	\centering
	\includegraphics[width=\textwidth]{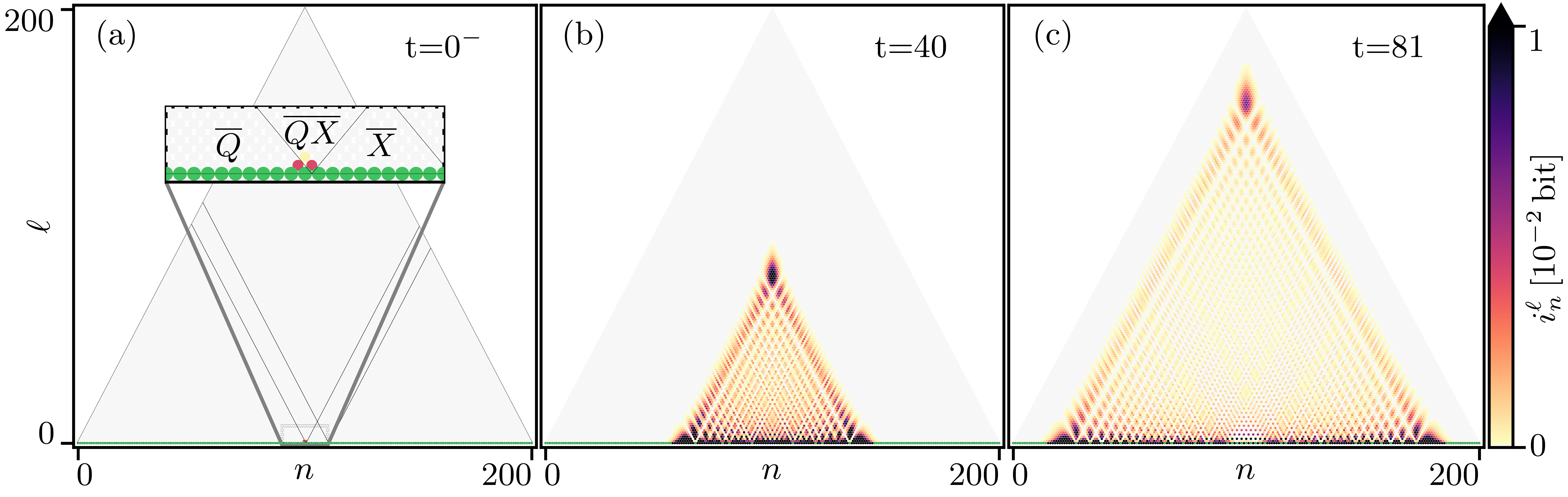}
	\caption{(a)-(c) Time evolution of local information after releasing a particle at the central site $\eta = 100$ in a tight-binding chain with $N=201$ sites.
    At $t=0^-$, the system is in the ground state of the Hamiltonian~\eqref{eq:HamTightBinding} with $\mu_i = - 20 \tau_p$ and $\mu_p = 20 \tau_p$.
    The post-quench dynamics is governed by the Hamiltonian~\eqref{eq:HamTightBinding} with $\mu_f = \mu_p = 20 \tau_p$.
    Sites with local information $1 \pm 0.01$ bits are marked in green.
    The thin black lines in (a) show the partitioning of the information lattice and
    the inset zooms into the information lattice near $\eta$ at $t=0^-$.
    Throughout the time evolution, the total information of the full information lattice remains conserved, that is, $I(\rho)=\sum_{n,l=0}^{N-1}i^\ell_n = 201$.
    }
	\label{fig:il_TB_SingleSite_NonCritical_SummaryPlots}
\end{figure*}

As a first step in studying the local information flow in quantum quench dynamics, we consider a simple protocol where one fermionic particle is released at the center of an empty tight-binding chain.
The system is governed by the tight-binding Hamiltonian
\begin{align}
    \nonumber
    H_\mathrm{TB}(\mu) &= (\mu-\mu_p) c_\eta^\dagger c^{\phantom{\dagger}}_\eta \\
    &+ \sum_{i=0}^{N-1} \mu_p c_i^\dagger c^{\phantom{\dagger}}_i + \sum_{i=0}^{N-2} \left( \frac{\tau_p}{2} c_i^\dagger c^{\phantom{\dagger}}_{i+1} +\text{H.c.} \right),
    \label{eq:HamTightBinding}
\end{align}
where $c_i^\dagger$ ($c_i$) creates (annihilates) spinless fermions, $N$ is the total number of sites in the physical chain, $\mu$ is the tunable chemical potential at the central site $\eta = \lfloor (N-1)/2 \rfloor$, and $\mu_p$ and $\tau_p$ are the chemical potential and the hopping amplitude of the rest of the chain.
By tuning $\mu$, we create a potential well ($\mu < \mu_p$) or a potential barrier ($\mu > \mu_p$) at site $\eta$.
At time $t=0$, we perform a local quench by suddenly changing $\mu$ from $\mu_i$ to $\mu_f$.
Specifically, at $t=0^-$, the system is prepared in the ground state $|\psi_0\rangle$ of the initial Hamiltonian $H_\mathrm{TB}(\mu_i)$ with $\mu_i = -20\tau_p$ and $\mu_p = 20\tau_p$.
For $t \geq 0$, the system evolves under the final Hamiltonian $H_\mathrm{TB}(\mu_f)$ with $\mu_f = \mu_p = 20\tau_p$, which is homogeneous.
Thus, the initial Hamiltonian $H_\mathrm{TB}(\mu_i)$ includes a potential well at $\eta$; its ground state $\lvert \psi_0 \rangle$ contains only one particle localized at $\eta$ while the rest of the chain is empty.
Due to the finite depth of the potential well, the wave function of the only particle present in the system has a small weight on the sites adjacent to $\eta$. 
This is reflected in the occupation-density distribution shown in Fig.~\ref{fig:occupation_density_NonCrit_TimeEvo}, in which at $t=0^-$ the central site is approximately fully occupied and all other sites are empty.

Turning to the information lattice, in the initial state $| \psi_0 \rangle$ most of the local information is localized at the smallest scale $\ell = 0$, as illustrated in Fig.~\ref{fig:il_TB_SingleSite_NonCritical_SummaryPlots}(a) where the green color corresponds to $i^\ell_n \approx 1$.
This short-scale localization of local information is characteristic of local product-like states~\cite{artiaco2024universal}.
Near site $\eta$, a small amount of information extends to $\ell \approx 1$ due to the short-range correlations associated with the overlap of the single-particle wave function localized at $\eta$ on adjacent sites [inset of Fig.~\ref{fig:il_TB_SingleSite_NonCritical_SummaryPlots}(a)].
In the final Hamiltonian $H_\mathrm{TB}(\mu_f)$, all sites have the same chemical potential $\mu = \mu_f  \gg \mu_i$.
As a consequence, during the post-quench dynamics, the particle released at $\eta$ propagates throughout the empty tight-binding chain in both the left and right directions.
This propagation is captured by the flow of local information from short scales $\ell \approx 0$ throughout the information lattice, as shown in Figs.~\ref{fig:il_TB_SingleSite_NonCritical_SummaryPlots}(b)-(c).

The flow of local information occurs mainly in two distinct directions.
Horizontally, local information flows away from $(n, \ell) \approx (\eta, 0)$ symmetrically, forming fronts that travel towards the left and right boundaries of the information lattice.
These fronts drag oscillating and decaying tails of local information, which are primarily concentrated at small scales and correspond to short-range correlations.
Vertically, local information moves from smaller to larger scales, signifying the development of longer range correlations.
The vertical front peaks at $n \approx \eta$ and travels at the same speed as the horizontal fronts.
It indeed carries the increasingly longer range correlations that emerge between the two horizontally moving fronts. 
At intermediate scales, local information patterns emerge that connect short- and long-range correlation fronts, moving diagonally within the information lattice.
They arise from the tails of the horizontally propagating fronts.
As discussed in Sec.~\ref{sec:info-lattice-properties}, the flow of information within the information lattice is local.
This implies that both horizontal and vertical fronts propagate with finite speed $v_\mathrm{LB} \approx \tau_p^{-1}$ according to the Lieb-Robinson bounds~\cite{lieb1972the}.

To understand the local information dynamics in this simple quench protocol, we start by considering the time evolution of the local occupation density in Fig.~\ref{fig:occupation_density_NonCrit_TimeEvo}.
This shows that the initially localized wave function of the particle at the central site spreads over time throughout the physical chain.
This spreading is governed by the expansion coefficients of the localized wave function in terms of the eigenstates of the final Hamiltonian, which are standing waves, and leads to the particle delocalization along the physical chain.
The information lattice quantifies the total amount of correlations at the local level that develop during the delocalization process. 

\begin{figure}[t]
	\centering
	\includegraphics[width=\columnwidth]{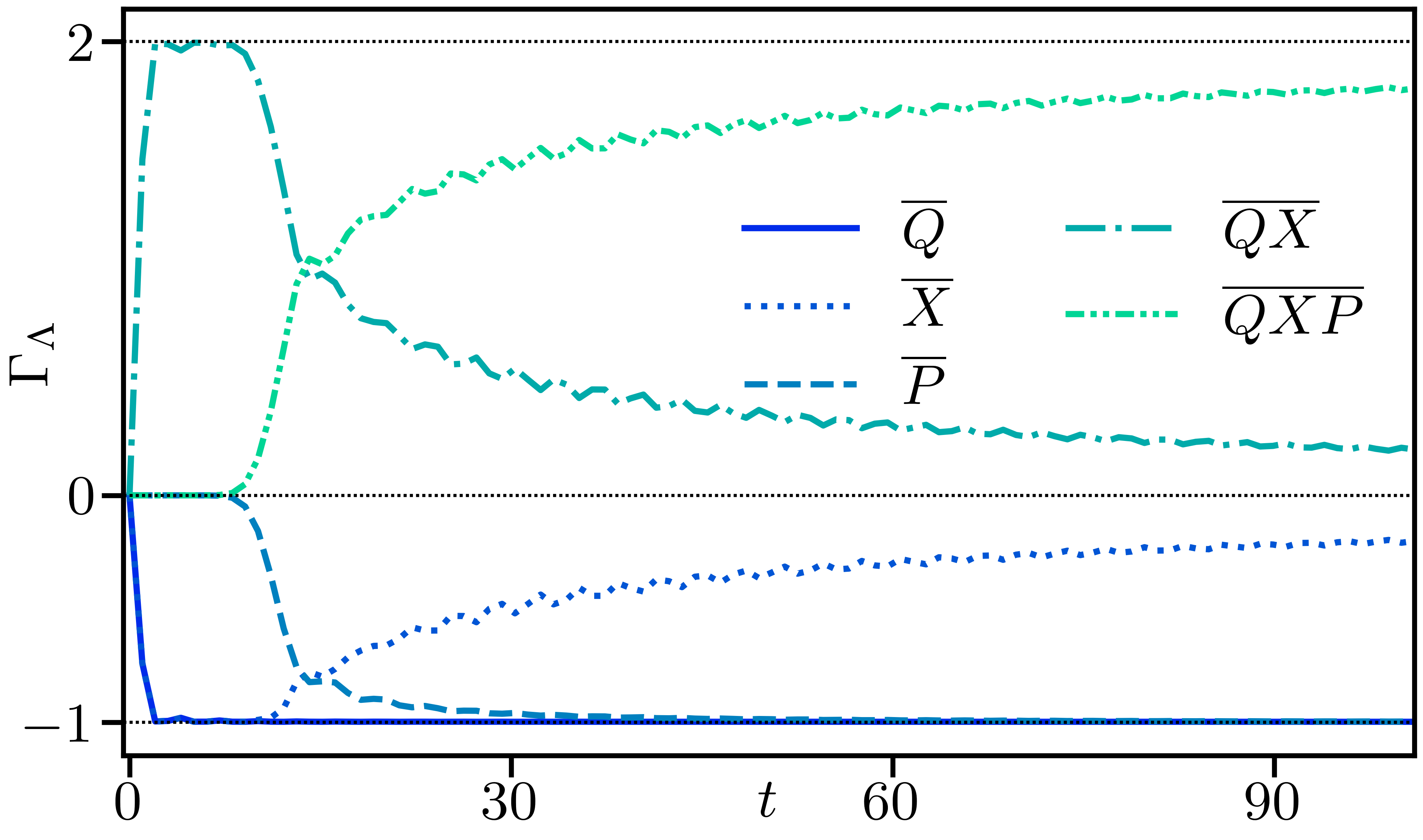}
	\caption{Change of total information $\Gamma_\Lambda$ in Eq.~\eqref{eq:totalInformation} for the chain decomposition in Fig.~\ref{fig:IL-Partition}: $Q$ (left half, $l_Q = 101$ including $\eta$), $X$ ($l_X = 10$) and $P$ ($l_P = 90$).
    The dynamics is induced by a quench protocol in which the system is prepared in the ground state of the Hamiltonian~\eqref{eq:HamTightBinding} with $\mu_i = -20 \tau_p$ and $\mu_p = 20 \tau_p$, and then evolved under the same Hamiltonian with $\mu_f = \mu_p = 20 \tau_p$.
    Horizontal dotted lines indicate expected asymptotic values of $\Gamma_\Lambda$ in each partition.}
	\label{fig:il_TB_SingleSite_NonCritical_SummaryPlots_totalInfo}
\end{figure}

Let us now turn to Fig.~\ref{fig:il_TB_SingleSite_NonCritical_SummaryPlots_totalInfo}, which illustrates the time behavior of $\Gamma_\Lambda$ in Eq.~\eqref{eq:totalInformation} for $l_Q=101$ (that is, $Q$ includes the left half of the chain with $\eta$ as the rightmost site), $l_X = 10$, and $l_P=90$.
As time progresses, local information shifts from the short-scale regions $\overline{Q}$ and $\overline{P}$ to the large-scale regions $\overline{QX}$ and $\overline{QXP}$.
The long-time values of $\Gamma_\Lambda$ approach quantized values of $-1, 0, +2$ bits~\footnote{Due to the presence of local information at intermediate scales connecting short- and long-range fronts, in a finite-size system the convergence to these quantized asymptotic values is not perfect.}.
In particular, the change of total information in $\overline{Q}$ and $\overline{P}$ converges to $-1$ bit.
At larger scales (first in $\overline{QX}$ and then in $\overline{QXP}$) it converges to $2$ bits, which corresponds to the local information shared between two maximally entangled systems with local Hilbert space dimension $2$.
This behavior reflects the spread of the initially localized wave function throughout the system.
The two horizontally propagating fronts evenly distribute the probability of finding the particle among the left and right halves of the chain.
Moreover, detecting the particle in either half determines its absence in the other one.
Thus, the left and right halves are in a maximally entangled state: $1/\sqrt{2} \left( | 0 \rangle_\mathrm{L} | 1 \rangle_\mathrm{R} + e^{i \theta} | 1 \rangle_\mathrm{L} | 0 \rangle_\mathrm{R}  \right)$ with $\theta$ the relative phase difference. 
The negative values of $\Gamma_\Lambda$ in $\overline{Q}$ and $\overline{P}$ may seem counterintuitive as information is transported into these regions via the horizontally moving fronts.
However, the scale resolution of the information lattice clarifies this effect: while a positive information packet propagates toward $\overline{P}$, local information at $ \ell \approx 0 $ decreases, leading to a net negative change in total information.
The reduction of the information corresponds to an increase of the von Neumann entropy of $P$ with the rest of the system.
Finally, as $l_X \ll l_{P/Q}$, $\Gamma_{\overline{X}}$ tends asymptotically to zero, which is the equilibrium value of the total information change in a small subsystem when the particle is delocalized over an infinitely large system.

\section{Quenching a single site in a critical tight-binding chain}
\label{sec:critical-states-quench}

\begin{figure*}[tb]
	\centering
	\includegraphics[width=\textwidth]{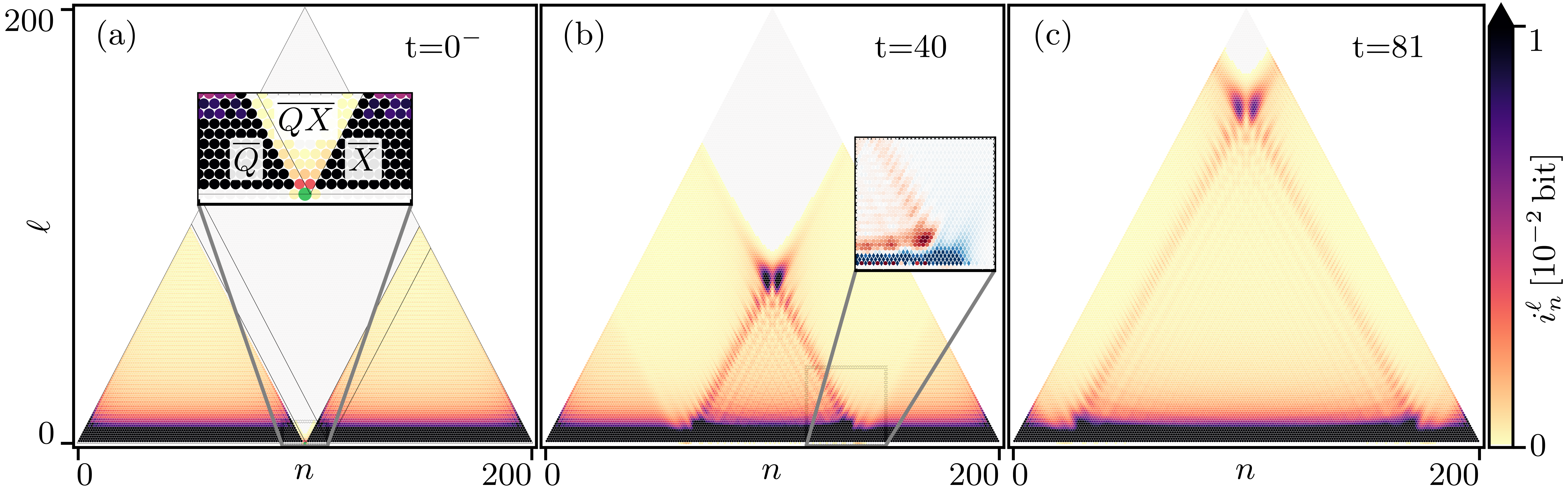}
	\caption{(a)-(c) Time evolution of local information after removing a potential barrier at the central site $\eta = 100$, which initially separates two tight-binding chains of $100$ sites each in a critical state.
    At $t=0^-$, the system is in the ground state of the Hamiltonian~\eqref{eq:HamTightBinding} with $\mu_i = 20 \tau_p$ and $\mu_p = 0$.
    The post-quench dynamics is governed by the Hamiltonian~\eqref{eq:HamTightBinding} with $\mu_f = \mu_p = 0$.
    Sites with local information $1 \pm 0.01$ bits are marked in green.
    The thin black lines in (a) show the partitioning of the information lattice and
    the inset zooms into the information lattice near $\eta$ at $t=0^-$.
    The inset in (b) shows the local-information change relative to the initial state [$\Delta i_n^\ell(t) = i_n^\ell(t) - i_n^\ell(0^-)$] at \(t = 40 \) with red dots (blue diamonds) indicating an increase (decrease) of local information.
    Darkest red correspond to $i^\ell_n \geq 0.01 $; darkest blue to $i^\ell_n \leq -0.01 $.
     Throughout the time evolution, the total information of the full information lattice remains conserved, that is, $I(\rho)=\sum_{n,l=0}^{N-1}i^\ell_n = 201$.
    }
	\label{fig:il_TB_SingleSite_Critical_SummaryPlots}
\end{figure*}

Connecting two initially isolated quantum systems is a fundamental quench protocol to probe transport and information dynamics.
Such a quench drives the system out of equilibrium and triggers the transport of particles or quasiparticles~\cite{piroli2017transport,alba2018entanglement}.
In this section, we investigate the local information flow generated by the sudden removal of a potential barrier at the center of a critical chain, effectively joining two previously disconnected critical halves~\cite{eisler2007evolution,calabrese2007entanglement,eisler2008entanglement,stephan2011local,eisler2014area,bernard2012energy,bhaseen2015energy,fischer2020energy}.

Specifically, at time $t=0^-$ the system is prepared in the ground state of the Hamiltonian~\eqref{eq:HamTightBinding} with the chemical potential at the central site $\eta = \lfloor (N-1)/2 \rfloor$ set to $\mu_i=20\tau_p$, while in the rest of the chain $\mu_p=0$~\footnote{To avoid numerical degeneracies, we use a small but finite $\mu_p = 10^{-5} \tau_p$ in simulations, while referring to $\mu_p = 0$ in the main text for conciseness.}.
For $t \geq 0$, the system evolves under the homogeneous Hamiltonian $H_\mathrm{TB}(\mu_f)$ with $\mu_f= \mu_p =0$.
Thus, before the quench, the system is composed of two critical tight-binding chains separated by a finite potential barrier at the central site.
The setup is reminiscent of a ``break junction" \cite{moreland1985electron,agrait2003quantum}.
To analyze the initial ground state $| \psi_0 \rangle$ of $H_\mathrm{TB}(\mu_i)$, it is instructive to first consider the limit of an infinite chemical potential at $\eta$, that is, $\mu_i \to \infty$.
In this case, the two critical chains are completely decoupled and $| \psi_0 \rangle$ is the product of their two ground states and the empty state at the potential barrier site: $ |\psi_0\rangle = |\psi_{0}^{\mathrm{half}}\rangle \otimes |\psi_{0,\eta}\rangle \otimes |\psi_{0}^{\mathrm{half}}\rangle$.
Here, $|\psi_{0}^{\mathrm{half}}\rangle = (\prod_{k \geq k_\mathrm{F}} c_k^\dagger) |0\rangle$ with $c_k^\dagger = \sqrt{\frac{2}{L+1}}\sum_{i=1}^{L} \sin(ki)\, c_i^\dagger$, where $L$ is the length of the tight-binding chain, and $k_\mathrm{F} = \pi/2$ the Fermi wave number.
Given the dispersion relation of the tight-binding chain, $E_\mathrm{TB}(k_n) = \tau_p \cos(k_n)$ with $k_n = \pi n / (L + 1)$ and $n = 1, 2, \dots, L$, for $k_\mathrm{F} = \pi/2$ the Fermi energy is zero, and the ground state $\ket{\psi^{\mathrm{half}}_0}$ features half of the single-particle states---in particular, those with the lowest energies---occupied. 
Therefore, the system is at half filling, critical, and has gapless excitations~\footnote{Notice that, strictly speaking, gapless excitations are only present in the thermodynamic limit.
The cost of the excitations decays linearly with system size.}.
When the potential barrier is finite, the two tight-binding chains to the left and right of $\eta$ are weakly coupled.
Then, the lowest energy single-particle state of the composed system is the symmetric linear combination of the lowest energy single-particle states of left and right chains; the second lowest energy single-particle state is the antisymmetric combination of the lowest energy single-particle states of left and right chains; and so on~\cite{jelic2012thedouble-well}.
The ground state $| \psi_0 \rangle$ features half of the single-particle states of the composite system---those with the lowest energies---occupied.
Thus, for a large but finite potential barrier, in $\ket{\psi_0}$ there is a small nonzero probability of finding a particle at the barrier site $\eta$.

Fig.~\ref{fig:il_TB_SingleSite_Critical_SummaryPlots}(a) illustrates the information lattice for the initial ground state $| \psi_0 \rangle$ of $H_\mathrm{TB}(\mu_i)$ with $\mu_i=20 \tau_p$ and $\mu_p=0$. 
To the left and right of $\eta$, two critical regions appear in which nonzero local information is present at all the information lattice sites of the triangles that extend up to $\ell = \lfloor (N-1)/2 \rfloor$.
Within these triangles, the average local information on scale $\ell$ decays as a power law, $\langle i^\ell_n \rangle \propto \ell^{-2}$, as implied by real-space scale invariance~\cite{artiaco2024universal}.
The inset of Fig.~\ref{fig:il_TB_SingleSite_Critical_SummaryPlots}(a) depicts a blow-up of the interface between the critical regions and the potential barrier, illustrating that nonzero local information at $\ell=0$ is present exclusively at $\eta$ while in the critical regions finite values of $i^\ell_n$ are only present for $\ell \geq 1$.
The green dot at $\eta$ signals that $i^0_\eta \approx 1$ bit.
This is a consequence of the barrier which approximately sets to zero the number of particles on $\eta$, implying that the outcome of the single-site operator $c^\dagger_\eta c^{\phantom{\dagger}}_\eta$ measuring the number of particles at $\eta$ has almost a certain outcome in $\ket{\psi_0}$.
However, the finite height of the potential barrier causes a small leakage of the single-particle wave function centered at $\eta$ on neighboring sites, as well as of the wave functions of the critical regions towards the barrier.
As a consequence, a small amount of local information ($i^\ell_n \lesssim 10^{-4}$) is present in the central region of the information lattice that extends in a light-cone shape from the site $\eta$.
With the chosen color scale, this effect is only visible at short scales along the diagonal interfaces between the critical triangles and the central region of the information lattice.
The absence of local information at $\ell=0$ within the critical regions is a consequence of them being approximately at half-filling.
The expectation value of the number operator at half-filling is $\langle c_i^\dagger c_i \rangle = 1/2$, which implies that the eigenvalues of $\rho_i$ are both $\lambda_{1,2} = 1/2$.
Thus, onsite reduced density matrices are fully mixed and have zero local information.

\begin{figure}[tbp]
	\centering
	\includegraphics[width=\columnwidth]{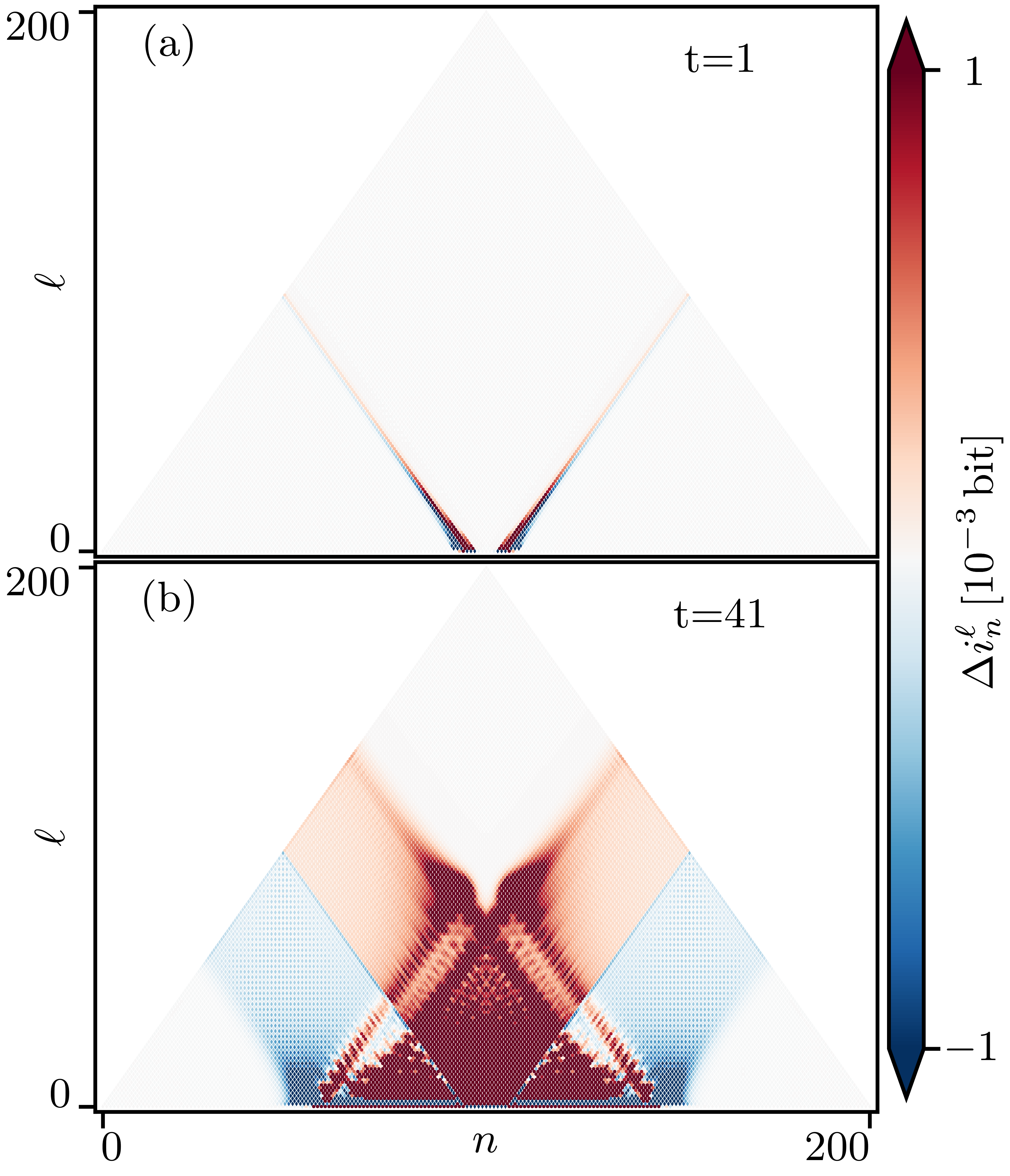}
	\caption{(a)-(b) Time evolution of the change in local information relative to the initial state, $\Delta i^\ell_n (t) = i^\ell_n(t) - i^\ell_n(0^-)$, at $t=1$ and $t=41$.
    Red dots (blue diamonds) indicate an increase (decrease) of local information with respect to the distribution at $t=0^-$.
    We perform the same quench protocol as in Fig.~\ref{fig:il_TB_SingleSite_Critical_SummaryPlots} but with a potential barrier in the pre-quench Hamiltonian spanning 11 physical sites centered at $\eta=100$.
    The barrier initially separates two critical chains.}
	\label{fig:il_TB_MultipleSite_Critical_SummaryPlots}
\end{figure}

Figs.~\ref{fig:il_TB_SingleSite_Critical_SummaryPlots}(b)-(c) show how local information flows through the information lattice after removing the potential barrier at $\eta$.
Recall that the post-quench dynamics takes place within a homogeneous ($\mu = \mu_p = 0$) critical system.
From the information lattice picture, the quench puts two reservoirs of local information in contact.
Similarly to Sec.~\ref{sec:potential-well-quench}, two main processes are triggered: horizontal flow of correlations and vertical flow of long-range correlations.
In this case, the horizontal flow occurs on top of the continuous information background characteristic of critical states.
This flow originates in the vicinity of the site $(n,\ell) = (\eta,0)$ and propagates information to the left and right boundaries of the information lattice.
The information propagating in one direction is composed of two fronts: one transporting a net positive information packet [red dots in the inset of Fig.~\ref{fig:il_TB_SingleSite_Critical_SummaryPlots}(b)] and another transporting a ``deficit" of information [blue dots in the inset of Fig.~\ref{fig:il_TB_SingleSite_Critical_SummaryPlots}(b)].
The net positive transport takes place at larger scales $\ell$ than the negative transport.
The inset also reveals that the deficit component slightly precedes the net positive one in spatial location $n$.
This decoupling between a positive and a negative component of the horizontal flow is more apparent for potential barriers that span more than one physical site.
Fig.~\ref{fig:il_TB_MultipleSite_Critical_SummaryPlots} shows the change in local information relative to the initial state, $\Delta i^\ell_n(t) = i^\ell_n(t) - i^\ell_n(0^-)$, at times $t = 1$ and $t = 40$.
The same quench protocol as in Fig.~\ref{fig:il_TB_SingleSite_Critical_SummaryPlots} is performed, but with a potential barrier in the pre-quench Hamiltonian that spans 11 physical sites located at the center of the physical chain.
These plots together with the information-reservoir interpretation of the critical tight-binding regions help us to understand the post-quench processes.
Connecting the two reservoirs through the quench, we induce a flow of local information from the right to the left reservoir, resulting in a net positive inflow of information in the left reservoir (and vice versa). 
Due to the conservation of total information, this inflow is fueled by a reduction of local information in the right reservoir (and vice versa).

Short-range information fronts are coupled at intermediate scales to the long-range information fronts via oscillatory local information patterns along the diagonal, similar to the previous quench scenario in Sec.~\ref{sec:potential-well-quench}.
The long-range information front is centered around the quenched potential barrier at $n \approx \eta$.
One prominent feature is the presence of two separate peaks of information that flow vertically from shorter to larger scales.
Fig.~\ref{fig:il_TB_MultipleSite_Critical_SummaryPlots}(b) clearly shows that each peak originates from one of the interfaces where the potential barrier meets the critical regions.
The right (left) peak quantifies the total amount of correlations between the negative propagating front (that is, the outflow of local information) in the right (left) reservoir and the positive propagating front in the left (right) reservoir.
The horizontal separation of the long-range information peaks scales with the size of the potential barrier [compare Fig.~\ref{fig:il_TB_SingleSite_Critical_SummaryPlots}(b) with Fig.~\ref{fig:il_TB_MultipleSite_Critical_SummaryPlots}(b)].

Another key feature of the quench protocols in Figs.~\ref{fig:il_TB_SingleSite_Critical_SummaryPlots}-\ref{fig:il_TB_MultipleSite_Critical_SummaryPlots} is that local information flows toward larger scales not only through the central peaks originating from the interfaces between the potential barrier and critical regions but also directly from the critical triangles containing information at $t=0^-$.
This is evident in Fig.~\ref{fig:il_TB_MultipleSite_Critical_SummaryPlots}(a).
Since nonzero local information exists within the critical regions up to $\ell \leq \lfloor (N-1) / 2 \rfloor$ in the initial state and the quench affects the central sites where the barrier is located, local information can suddenly flow from any lattice site within a light cone originating from the quenched sites.
This is illustrated by the diagonal red and blue lines at the interface between the initial critical triangles and the central region of the information lattice.
The Lieb-Robinson bound fixes the maximal propagation velocity of information within the system, independent of the scale where the local information is located.
Hence, in the critical chain the quenched sites become immediately correlated with the farthest physical site in the critical region since correlations are initially present at all scales.

\begin{figure}[tbp]
	\centering
	\includegraphics[width=\columnwidth]{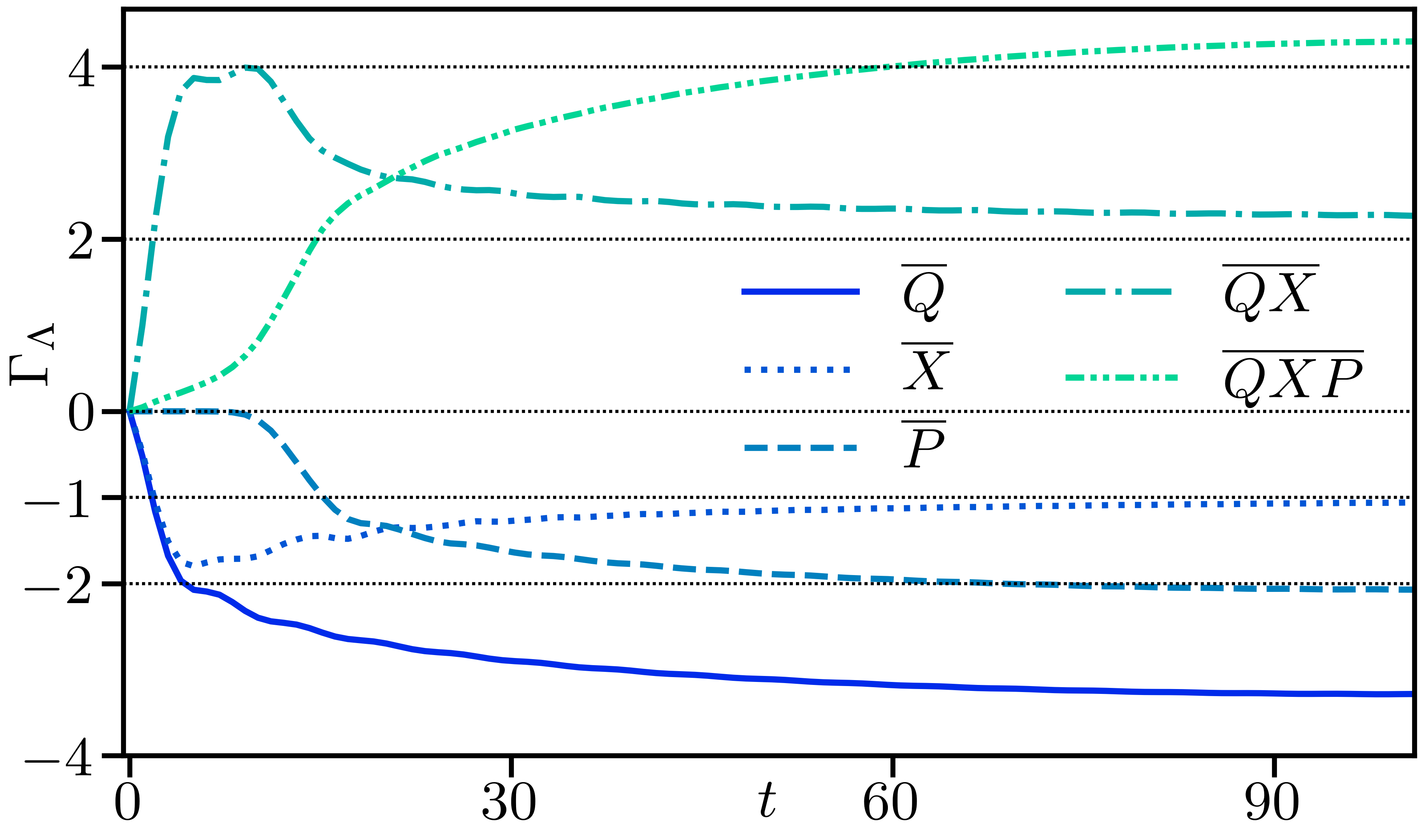}
	\caption{Change of total information $\Gamma_\Lambda$ in Eq.~\eqref{eq:totalInformation} for the chain decomposition in Fig.~\ref{fig:IL-Partition}: $Q$ (left half, $l_Q = 101$ including $\eta$), $X$ ($l_X = 10$) and $P$ ($l_P = 90$).
    The dynamics is induced by a quench protocol in which the system is prepared in the ground state of the Hamiltonian~\eqref{eq:HamTightBinding} with $\mu_i = 20 \tau_p$ and $\mu_p = 0$, and then evolved under the same Hamiltonian with $\mu_f = \mu_p = 0$.
    Horizontal dotted lines are guides to the eye set to $-2, -1, 2, 4$ bits.}
	\label{fig:il_TB_SingleSite_Critical_TotalInfoPlots}
\end{figure}

Fig.~\ref{fig:il_TB_SingleSite_Critical_TotalInfoPlots} shows the time dependence of $\Gamma_\Lambda$ in Eq.~\eqref{eq:totalInformation} for $l_Q=101$ (where $Q$ includes the left half of the chain with $\eta$ as the rightmost site), $l_X=10$ and $l_P=90$.
As in the previous quench scenario in Sec.~\ref{sec:potential-well-quench}, we observe a flow of local information toward larger-scale partitions $\overline{QX}$ and $\overline{QXP}$, while information within the smaller-scale partitions $\overline{Q}$, $\overline{X}$ and $\overline{P}$ decreases.
The total-information change $\Gamma_\Lambda$ in $\overline{Q}$, $\overline{X}$ and $\overline{P}$ exhibits characteristic signatures observed in the time evolution of the von Neumann entropy following local bond-defect quenches in critical chains~\cite{eisler2007evolution,calabrese2007entanglement,eisler2008entanglement,stephan2011local}.
Specifically, the von Neumann entropy of a subsystem has a characteristic logarithmic behavior over time, reaches a maximum that scales logarithmically with subsystem size, and for finite-size subsystems within a sufficiently large system eventually decays to its equilibrium value~\cite{eisler2007evolution,calabrese2007entanglement}.
$\Gamma_\Lambda$ within $\overline{Q}$ and $\overline{P}$ follows a log-scale decrease, while $\Gamma_{\overline{X}}$ seems to converge to a negative finite value.
Throughout the time evolution, $\Gamma_\Lambda$ in $\overline{Q}$, $\overline{X}$ and $\overline{P}$ exhibit step-like oscillations due to finite-size effects and the nonlinear dispersion~\cite{eisler2007evolution,eisler2008entanglement}.
For the larger-scale partitions $\overline{QX}$ and $\overline{QXP}$, we observe similar trends in the change of total information.
$\Gamma_{\overline{QX}}$ follows a logarithmic increase, briefly stabilizes in a plateau regime, and then decays, approaching a finite value.
The concurrent flow of local information at all scales and the associated logarithmic scaling of the total information change are the hallmarks of critical states.

\section{Connecting a topological Kitaev chain to a critical tight-binding chain}
\label{sec:majorana-teleportation-quench}

\begin{figure*}[tb]
 	\centering
 	\includegraphics[width=\textwidth]{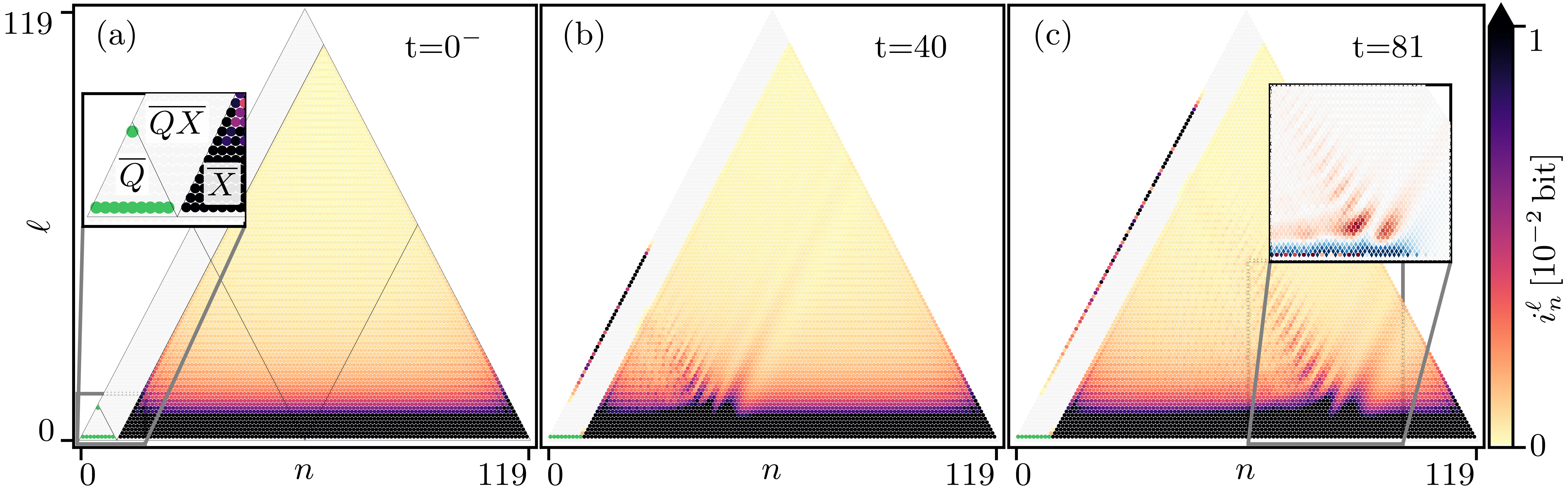}
 	\caption{(a)-(c) Time evolution of local information after coupling a topological sweet-spot Kitaev chain in the $Q$ region to a tight-binding chain in the $XP$ region.
    At $t=0^-$, the system is in the product state of the ground states of the Hamiltonians~\eqref{eq:HamKitaevChainAtTSS} and \eqref{eq:HamQuenchProbeTB} with $\tau_t = 0$, $\tau = 20 \tau_p$, and $\tau_p = 1$.
    The post-quench dynamics is governed by the Hamiltonians~\eqref{eq:HamKitaevChainAtTSS} and \eqref{eq:HamQuenchProbeTB} with $\tau_t = \tau_p$, $\tau = 20 \tau_p$, and $\tau_p = 1$.
    The length of the Kitaev chain is $l_Q=10$ and of the tight-binding chain is $l_X+l_P=110$.
    Sites with local information $1 \pm 0.01$ bits are marked in green.
    The thin black lines in (a) show the partitioning of the information lattice and the inset zooms into the information lattice in partition $\overline{Q}$ at $t=0^-$.
    The inset in (c) shows changes in local information relative to the initial state [$\Delta i_n^\ell(t) = i_n^\ell(t) - i_n^\ell(0^-)$] at $t = 81$ with red dots (blue diamonds) indicating an increase (decrease) of local information.
    Darkest red correspond to $i^\ell_n \geq 0.01 $; darkest blue to $i^\ell_n \leq -0.01 $.
     Throughout the time evolution, the total information of the full information lattice remains conserved, that is, $I(\rho)=\sum_{n,l=0}^{N-1}i^\ell_n = 120$.
    }
 	\label{fig:il_MZM_teleportation_SummaryPlots}
\end{figure*}

As a final quench protocol, we consider suddenly connecting a topological Kitaev chain to a critical tight-binding chain.
This setup is inspired by Ref.~\cite{bauer2023quench}.
The Kitaev chain forms the $Q$ region in Fig.~\ref{fig:IL-Partition}(c), extending over $l_Q$ physical sites.
It is governed by the sweet-spot Hamiltonian
\begin{align}
	H^Q_{\mathrm{KC}}&= \frac{\tau}{2}\sum_{i=0}^{l_Q-2} \left( c_i^\dagger c^{\phantom{\dagger}}_{i+1} +  c^{\phantom{\dagger}}_i c^{\phantom{\dagger}}_{i+1}+\text{H.c.} \right) \nonumber \\
    &= i\frac{\tau}{2} \sum_{i=0}^{l_Q-2} \gamma_{B,i}\gamma_{A,i+1} =\tau\sum_{i=0}^{l_Q-2} \left(d_i^\dagger d^{\phantom{\dagger}}_i - \frac{1}{2}\right).
	\label{eq:HamKitaevChainAtTSS}
\end{align}
This Hamiltonian is diagonalized by expressing the original fermions in terms of Majorana operators $c_i = \frac{1}{2}\left(\gamma_{B,i} + i\gamma_{A,i}\right)$ with $\gamma^{\phantom{\dagger}}_{\alpha,i} = \gamma_{\alpha,i}^\dagger$ and $\{\gamma_{\alpha,i},\gamma_{\beta,j}\}=2\delta_{\alpha \beta}\delta_{ij}$, and introducing the fermionic operators $d_i= \frac{1}{2}\left(\gamma_{B,i}+i\gamma_{A,i+1}\right)$ that pair Majorana operators on neighboring sites $[i,i+1]$.
The left- ($\gamma_L := \gamma_{A,0}$) and right- ($ \gamma_R := \gamma_{B,l_Q-1}$) edge Majorana operators are absent in Eq.~\eqref{eq:HamKitaevChainAtTSS}; they together form a delocalized fermionic mode: $f=\frac{1}{2}\left(\gamma_R+i\gamma_L\right)$ and $f^\dagger =\frac{1}{2}\left(\gamma_R-i\gamma_L\right)$.
Occupying the delocalized fermionic mode costs zero energy~\cite{alicea2012new}.
Thus, the eigenstates of the sweet-spot Kitaev chain are twofold degenerate.
The bulk modes $d_i$ exhibit a flat energy dispersion $E^Q_\mathrm{KC} = \tau$.

The tight-binding chain forms the $XP$ region in Fig.~\ref{fig:IL-Partition}(c) and is governed by the Hamiltonian
\begin{align}
\nonumber
    H_{\mathrm{TB}}^{XP}(\tau_t) &= \frac{\tau_t}{2} \left(c_{l_Q-1}^\dagger c^{\phantom{\dagger}}_{l_Q} + \text{H.c.} \right)\\
    &+ \sum_{i=l_Q}^{N-2} \frac{\tau_p}{2} (c^\dagger_i c^{\phantom{\dagger}}_{i+1} + \text{H.c.} ),
    \label{eq:HamQuenchProbeTB}
\end{align} 
where $\tau_p$ is the intra-chain hopping amplitude and the coupling to the Kitaev chain is achieved through a tunneling term of strength $\tau_t$.
$N$ is the total number of sites in the physical chain composed of both the Kitaev chain and the tight-binding chain.
For $\tau_t = 0$, the spectrum of the tight-binding Hamiltonian is $E^{XP}_{\mathrm{TB}}(k)=\tau_p \cos(k_n)$ (see Sec.~\ref{sec:critical-states-quench}).

At $t = 0^-$, the system is prepared in the ground state $|\psi_{0}\rangle$ of the Hamiltonians~\eqref{eq:HamKitaevChainAtTSS} and \eqref{eq:HamQuenchProbeTB} with $\tau_t =0$ and $\tau = 20 \tau_p$.
Thus, $|\psi_{0}\rangle$ is the product state of the ground state of the sweet-spot Kitaev chain $|\psi_{0,\mathrm{KC}}\rangle$, which verifies $f |\psi_{0,\mathrm{KC}}\rangle = d_i |\psi_{0,\mathrm{KC}}\rangle = 0$ for all $d_i$, and the critical ground state of the tight-binding chain at half filling $|\psi_{0,\mathrm{TB}}\rangle$ described in Sec.~\ref{sec:critical-states-quench}.

Fig.~\ref{fig:il_MZM_teleportation_SummaryPlots}(a) depicts the information lattice for the initial state $|\psi_{0}\rangle$.
The topological ground state $|\psi_{0,\mathrm{KC}}\rangle$ has 1 bit of local information at the top of partition $\overline{Q}$, $i^{l_Q-1}_{(l_Q-1)/2} = 1$.
Such a $\mathcal{O}(1)$ contribution to the information at system size scales is the hallmark of topological states~\cite{artiaco2024universal}.
For $|\psi_{0,\mathrm{KC}}\rangle$, $i^{l_Q - 1}_{(l_Q - 1)/2} = 1$ implies the existence of a two-outcome measurement of the correlations between the edge sites $0$ and $l_Q-1$ that can be predicted with certainty.
Specifically, the measurement amounts to determining the eigenvalue of $n_f = f^\dagger f$---the occupation of the delocalized fermionic mode.
$|\psi_{0,\mathrm{KC}}\rangle$ also has an extensive amount of local information at scale $\ell=1$, where 1 bit is shared between every neighboring physical site, reflecting the ``dimer" structure of the bulk of the sweet-spot Kitaev chain apparent from the last term in Eq.~\eqref{eq:HamKitaevChainAtTSS}.
Local information in the critical ground state of the tight-binding chain presents the power-law decay $\langle i^\ell_n \rangle \propto \ell^{-2}$ discussed in Sec.~\ref{sec:critical-states-quench}.
Of particular interest for this quench is the information lattice interface $\widetilde{QXP}$ illustrated in Fig.~\ref{fig:IL-Partition}(c), which contains local information values $\{ i_{l_Q - 1 + \ell/2}^\ell \}$ with $ \ell = 1,\dots,N - l_Q + 1$ on the diagonal line that starts at $(n, \ell) = (l_Q - 1/2, 1)$.
At $t=0^-$, there is no information in $\widetilde{QXP}$, as shown in Fig.~\ref{fig:il_MZM_teleportation_SummaryPlots}(a), indicating that, in the initial state, there are no correlations involving the rightmost physical site of the Kitaev chain $l_Q-1$ and the $XP$ region.

\begin{figure}[t]
 	\centering
 	\includegraphics[width=\columnwidth]{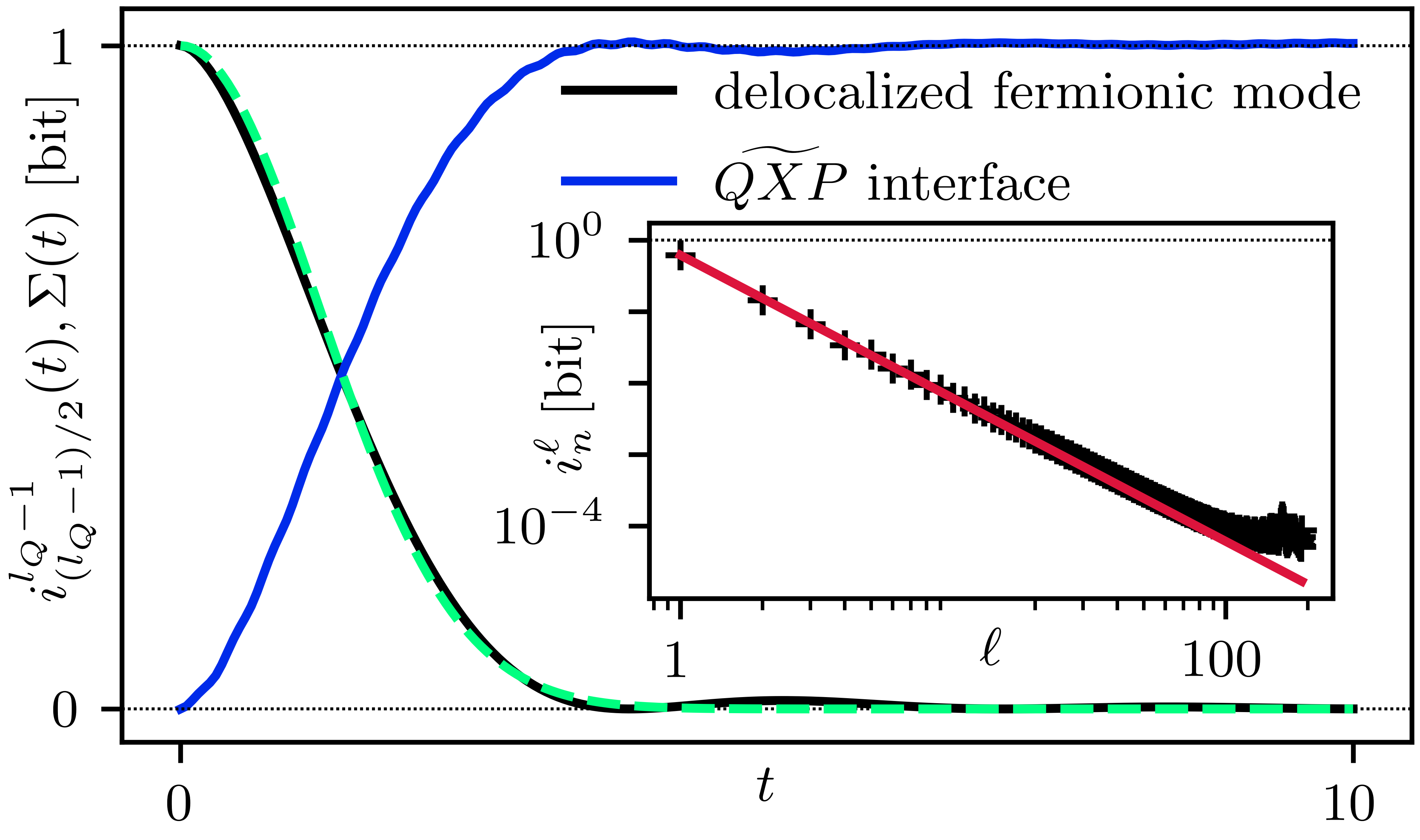}
 	\caption{Time evolution of $i^{l_Q-1}_{(l_Q-1)/2}(t)$ (black), which is the information linked to the delocalized fermionic mode occupation $n_f = f^\dagger f$, and $\Sigma(t) = \sum_{(n,\ell) \in \widetilde{QXP}} i^\ell_n(t)$ (blue), which is the total information in the $\widetilde{QXP}$ interface, for the same quench protocol as in Fig.~\ref{fig:il_MZM_teleportation_SummaryPlots}.
    An exponential fit $i_{(l_Q-1)/2}^{l_Q-1}(t)\propto\exp(-\alpha t^2)$ (green dashed, with $\alpha \approx 0.36$) matches analytical predictions from Majorana correlation functions\protect~\cite{goldstein2011decay,budich2012failure}.
    Horizontal dotted lines mark expected asymptotic values.
    The inset shows local information across $\widetilde{QXP}$ at $t \approx 172$ (black crosses), showing power-law decay $i^\ell_{l_Q-1+\ell/2} \propto \ell^{-2}$ (red), for the same quench protocol as in Fig.~\ref{fig:il_MZM_teleportation_SummaryPlots} but in a larger system with $l_Q = 10$ and $l_X + l_P = 190$.
	}
 	\label{fig:QX-Interface_TotalInfoPlots}
\end{figure}

For $t \geq 0$, the system evolves under the Hamiltonians~\eqref{eq:HamKitaevChainAtTSS} and \eqref{eq:HamQuenchProbeTB} with intra-chain parameters $\tau$ and $\tau_p$ identical to the pre-quench configuration ($\tau = 20 \tau_p$) and the inter-chain tunneling strength set to $\tau_t = \tau_p$.
Figs.~\ref{fig:il_MZM_teleportation_SummaryPlots}(b)–(c) show two snapshots of the time-evolved local information.
The most prominent feature is the clear separation between the local information flow originating from the top of $\overline{Q}$ and the flow across the rest of the information lattice.
This decoupling implies that the topological information of the Kitaev chain remains well-defined and quantized to 1 bit. 
This topological information flows towards larger scales along the diagonal left boundary of the information lattice, confined to a single site in the horizontal direction while exhibiting a long oscillatory tail diagonally. 
Due to this decoupling, the sum of local information along the diagonal starting at $(n,\ell)=(\frac{l_Q-1}{2},l_Q-1)$ and ending at the top of the information lattice remains quantized to 1 bit.
Local information at the top of $\overline{Q}$, $i^{l_Q-1}_{(l_Q-1)/2}(t)$, decays over time with a quadratic exponential form as illustrated by the black curve in Fig.~\ref{fig:QX-Interface_TotalInfoPlots}, in agreement with analytical predictions for Majorana correlations and their lifetimes in similar setups~\cite{goldstein2011decay,budich2012failure}.

The quench also induces a horizontally propagating information packet that travels across the critical region on top of the local information reservoir toward the right boundary of the information lattice.
Over time, this right-traveling packet slowly spreads, gradually smearing information across the entire critical region.
Simultaneously, local information quickly flows from the partitions $\overline{X}$ and $\overline{XP}$ to the $\widetilde{QXP}$ interface. 
The presence of such a fast multiscale transfer is the hallmark of critical-state dynamics and resembles the behavior observed in Fig.~\ref{fig:il_TB_MultipleSite_Critical_SummaryPlots}.
This transfer stops when the total information in $\widetilde{QXP}$ converges to 1 bit, as illustrated by the blue curve in Fig.~\ref{fig:QX-Interface_TotalInfoPlots}.
The asymptotic distribution of local information in $\widetilde{QXP}$ has a power-law decay $i^\ell_n \propto \ell^{-2}$, resembling the pre-quench distribution in the critical state of the tight-binding chain (see inset).

An analytical understanding of the above behaviors begins by noting that, to leading order in the limit $\tau \gg \tau_{p/t}$, the bulk degrees of freedom in $Q$ are frozen and do not contribute to the dynamics; that is, the operators $d_i^\dagger d_i$ all have fixed eigenvalues.
Consequently, local information in partitions $ \overline{Q}$, $\overline{QX}$, and $\overline{QXP}$ remains fixed at its initial value, except at the left and right edges.
Specifically, $i^1_n = 1$ for $n = 1/2, \dotsc, l_Q - 3/2$, while all other values away from the edges vanish.
Formally, this follows from adiabatic elimination, with $\tau^{-1}$ as the small parameter, whose leading-order result is the Hamiltonian $H^{QXP} = H^{Q} + H^{XP}$ projected onto the ground state manifold of $H^{Q}$
\begin{align}
    \nonumber
    H_{\text{eff}}^{QXP} &= \frac{\tau_t}{4} \left(c_{l_Q}^\dagger - c^{\phantom{\dagger}}_{l_Q} \right)(f+f^\dagger)\\
    &+ \sum_{i=l_Q}^{N-2} \frac{\tau_p}{2} (c^\dagger_i c^{\phantom{\dagger}}_{i+1} + \text{H.c.} ) + \mathcal{O}\left(\frac{\tau_{t/p}^2}{\tau}\right).
    \label{eq:effHamQuenchProbeTB}
\end{align}
The higher orders in the perturbation theory $\mathcal{O}\!\left(\tau_{t/p}^2/\tau\right)$ account for couplings between the site $l_Q - 2$ and the tight-binding chain.
Thus, effectively, $Q$ is made of a single lattice site---with an associated creation (annihilation) operator $f^\dagger$ ($f$) shared between the edges of $Q$---coupled to the tight-binding chain.

\begin{figure}[t]
 	\centering
 	\includegraphics[width=\columnwidth]{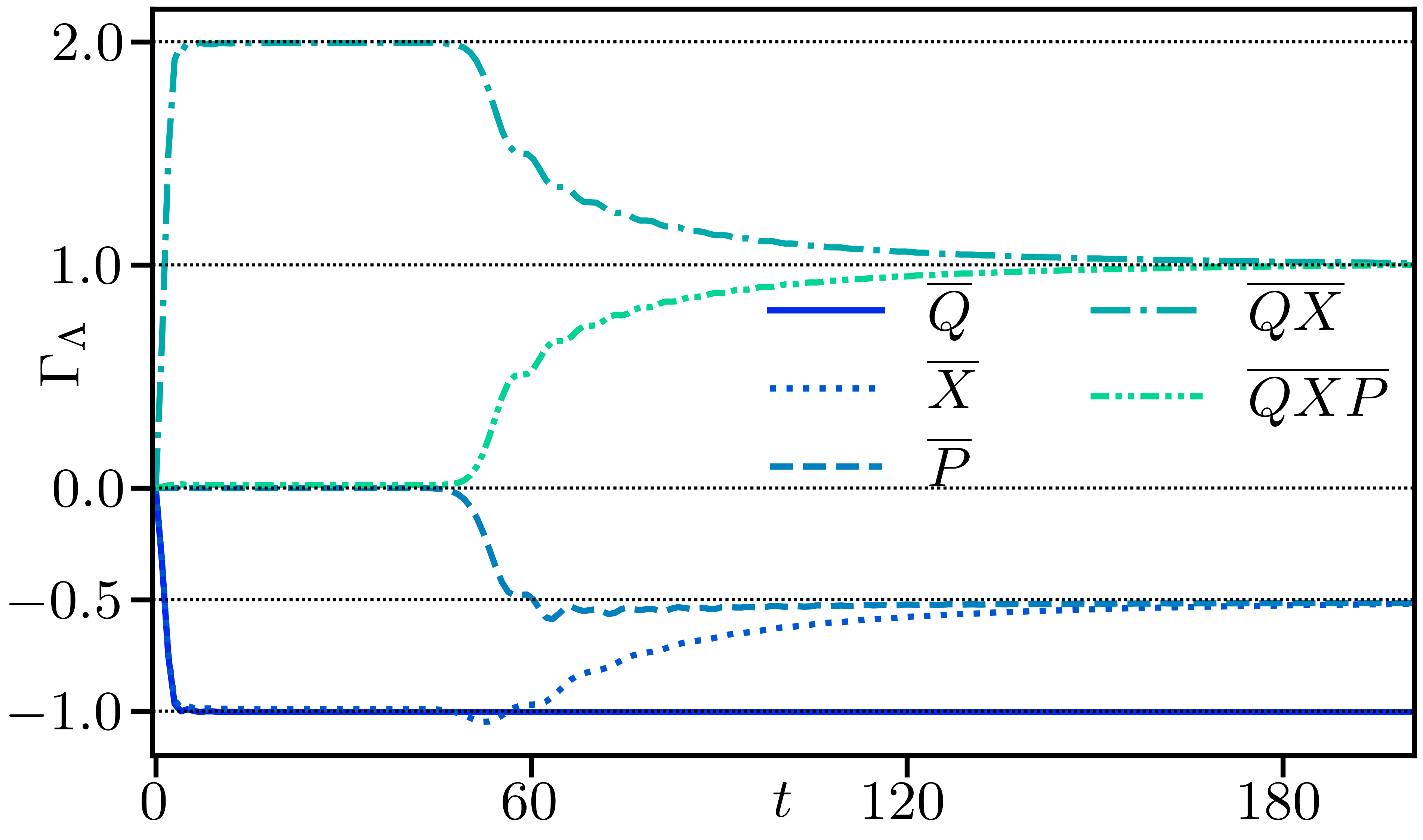}
 	\caption{Change of total information $\Gamma_\Lambda$ in Eq.~\eqref{eq:totalInformation} for the chain decomposition in Fig.~\ref{fig:IL-Partition}: $Q$ (left half, $l_Q = 10$), $X$ ($l_X = 50$) and $P$ ($l_P = 140$).
    The dynamics is induced by a quench protocol in which the system is prepared in the product state of the ground states of the Hamiltonians~\eqref{eq:HamKitaevChainAtTSS} and \eqref{eq:HamQuenchProbeTB} with $\tau_t = 0$, $\tau = 20 \tau_p$, and $\tau_p = 1$, and then evolved under the same Hamiltonian with $\tau_t = \tau_p$, $\tau = 20 \tau_p$, and $\tau_p = 1$.
    Horizontal dotted lines indicate expected asymptotic values of $\Gamma_\Lambda$ in each partition.
    }
    \label{fig:il_MZM_teleportation_TotalInfoPlots}
\end{figure}

Let us start by considering the short-time behavior in Figs.~\ref{fig:il_MZM_teleportation_SummaryPlots}, \ref{fig:QX-Interface_TotalInfoPlots}, and \ref{fig:il_MZM_teleportation_TotalInfoPlots}, that is, the rapid loss of 1 bit of information in $\overline{Q}$ flowing to larger scales from the top of this partition, accompanied by the simultaneous loss of 1 bit in $\overline{X}$ and the gain of 2 bits in $\overline{QX}$.
This is particularly clear in Fig.~\ref{fig:il_MZM_teleportation_TotalInfoPlots} illustrating the time dependence of $\Gamma_\Lambda$ in Eq.~\eqref{eq:totalInformation} for $\Lambda \in \{ \overline{Q}, \overline{X}, \overline{P}, \overline{QX}, \overline{QXP} \}$ and $l_Q=10$, $l_X=50$ and $l_P=140$.
We notice that the effective model in Eq.~\eqref{eq:effHamQuenchProbeTB} is critical and exhibits particle-hole symmetry under the transformation $f \leftrightarrow f^\dagger$.  
Thus, for $t \gtrsim \tau_{t/p}^{-1}$ the time-evolved state locally respects particle-hole symmetry~\cite{calabrese2006time}, implying $\braket{n_f} = 1/2$.  
This means that the state of the effective single site in the $Q$ region becomes maximally mixed.  
Since this state was initially pure, this implies that $\overline{Q}$ loses 1 bit of information.  
Moreover, since the full system is pure, the von Neumann entropy of the $XP$ region must necessarily equal that of the $Q$ region.  
Consequently, as the state in $XP$ was initially pure, the union partition $\overline{X}\cup\overline{P}\cup\overline{XP}$ must also lose exactly 1 bit of information.  
Due to local conservation of information, these 2 bits (one from $\overline{Q}$ and one from $\overline{X}\cup\overline{P}\cup\overline{XP}$) cannot go anywhere other than in the $\overline{QX}$ partition.  
The approximate time it takes before any changes occur in the $\overline{QXP}$ partition is given by the Lieb-Robinson speed $v_\mathrm{LB} \approx \tau_p^{-1}$ multiplied by the length of the $X$ region~\cite{lieb1972the,klein2022time}, that is, $t \approx v_\mathrm{LB} l_X \approx \tau_p^{-1} l_X$.  
In conclusion, the above arguments imply that for times satisfying $\tau_p^{-1} \lesssim t \lesssim \tau_p^{-1} l_X$, the partitions $\overline{Q}$ and $\overline{X}$ each have roughly 1 bit less information, while the $\overline{QX}$ partition gains 2 bits compared to the initial state, in agreement with the observed data in Figs.~\ref{fig:il_MZM_teleportation_SummaryPlots}, \ref{fig:QX-Interface_TotalInfoPlots}, and \ref{fig:il_MZM_teleportation_TotalInfoPlots}.

We now turn to the analytical understanding of the late-time behavior shown in Figs.~\ref{fig:il_MZM_teleportation_SummaryPlots}, \ref{fig:QX-Interface_TotalInfoPlots}, and \ref{fig:il_MZM_teleportation_TotalInfoPlots}.
For times $t \gtrsim \tau_p^{-1} l_X$, $\Gamma_{\overline{Q}}$ saturates to $-1$ bit, $\Gamma_{\overline{X}}$ and $\Gamma_{\overline{P}}$ each saturate to $-1/2$ bit, $\Gamma_{\overline{QX}}$ converges asymptotically to $1$ bit located at the $\widetilde{QXP}$ interface and distributed according to $i^\ell_{l_Q-1+\ell/2} \propto \ell^{-2}$, and $\Gamma_{\overline{QXP}}$ saturates to $1$ bit that flows in this partition from the top of $\overline{Q}$ where the topological edge correlation initially resides.  
As shown earlier, 1 bit is lost from both $\overline{Q}$ and $\overline{XP}$ and transferred into correlations between $Q$ and $XP$. 
What remains to be explained is why, at long times, the bit lost from $\overline{X} \cup \overline{XP} \cup \overline{P}$ is equally split between $\overline{X}$ and $\overline{P}$, and why the 2 gained bits are equally distributed between $\overline{QX}$ and $\overline{QXP}$.

To address these questions, we note that the energy of the time-evolved state exceeds that of the ground state of the Hamiltonian~\eqref{eq:effHamQuenchProbeTB} by at most $2\tau_t$---the difference between the largest and smallest eigenvalues of the first term in the Hamiltonian.
Since the system is an energy conductor, this excess in energy density initially located at the boundary between $Q$ and $XP$ dissipates over time and, in the long-time limit, the reduced density matrices of small subsystems converge to those of the ground state, assuming a sufficiently large system.
In the quench protocol of Fig.~\ref{fig:il_MZM_teleportation_TotalInfoPlots}, where $l_X \ll l_P$, the long-time reduced density matrices within the $QX$ region thus approach those of the ground state.
If the effective Hamiltonian~\eqref{eq:effHamQuenchProbeTB} was nondegenerate, its critical ground state would feature a power-law decay of local information $i^\ell_n \propto \ell^{-2}$, and for $l_X \gg 1$ the full 2 bits of correlation between $Q$ and $XP$ would be localized within $\overline{QX}$.
The same would then hold for the post-quench time-evolved state at late times.
However, $H_{\mathrm{eff}}^{QXP}$ is two-fold degenerate, which complicates this picture.
Since $H_{\mathrm{eff}}^{QXP}$ commutes with $f - f^\dagger$, any eigenstate $\ket{\phi}$ of $H_{\mathrm{eff}}^{QXP}$ with definite fermion parity has a degenerate partner $(f - f^\dagger)\ket{\phi}$ with opposite parity.
In the presence of such a two-fold degeneracy, the ground state may host 1 nonlocal bit of information~\cite{artiaco2024efficient}.  
Thus, only 1 of the 2 bits of correlation between $Q$ and $XP$ must lie in $\overline{QX}$; the remaining bit, tied to the degeneracy, may reside in $\overline{QXP}$.

To understand whether this bit resides in $\overline{QXP}$, we first note that the time-evolved state remains an eigenstate of the backward-in-time Heisenberg operator $n_f(t)=f^\dagger(t) f(t)$ at all times, where we define $f(t)=U(t)\,f\,U^\dagger(t)$ with $U(t)=\exp \bigl(-i H_{\mathrm{eff}}^{QXP} t\bigr)$.
The definite outcome of $n_f(t)$ implies the existence of one binary question that can always be answered with certainty, corresponding to 1 bit of information.
The quench causes $f(t)+f^\dagger(t) = \gamma_R(t)$ to spread into the critical region, similar to the single-particle spreading in Sec.~\ref{sec:potential-well-quench}.
For $l_P \gg l_X$, the operator $f(t)+f^\dagger(t)$ becomes almost entirely supported in $P$, while $f(t)-f^\dagger(t) = \gamma_L(t)$ remains localized in $Q$, as it commutes with the Hamiltonian.
Hence, the 1 bit of information associated with $n_f(t)$ becomes a correlation between $Q$ and $P$.  
Eventually, of the 2 bits of total correlation between $Q$ and the rest of the system, one resides in the $\widetilde{QXP}$ interface and the other in the long-range correlation between $Q$ and $P$, that is, in the $\overline{QXP}$ partition.
This implies that $\overline{X}$ and $\overline{P}$ must each lose half a bit of information, in agreement with Fig.~\ref{fig:il_MZM_teleportation_TotalInfoPlots} and with the static and dynamical fractional entanglement signature of Majorana zero modes previously reported in Refs.~\cite{sela2019detecting,bauer2023quench}.

\subsection{Deviation from the topological sweet spot}

The qualitative behavior observed in Figs.~\ref{fig:il_MZM_teleportation_SummaryPlots}, \ref{fig:QX-Interface_TotalInfoPlots}, and \ref{fig:il_MZM_teleportation_TotalInfoPlots} apply more generally than to the sweet-spot Kitaev Hamiltonian~\eqref{eq:HamKitaevChainAtTSS}.
Specifically, the following three previously discussed phenomena result solely from the validity of the effective Hamiltonian~\eqref{eq:effHamQuenchProbeTB}:
(i)~The flow of 1 bit of information associated with topological edge correlations towards larger scales along the diagonal left boundary of the information lattice is decoupled from the local information flow within the critical region, as shown in Fig.~\ref{fig:il_MZM_teleportation_SummaryPlots}(a).
(ii) At short times, partitions $\overline{Q}$ and $\overline{X}$ lose 1 bit of information each, while $\overline{QX}$ gains 2 bits of information, as shown in Fig.~\ref{fig:il_MZM_teleportation_TotalInfoPlots}.
(iii) At long times, the bit associated with topological edge correlations transfers to $\overline{QXP}$, indicating a net flow of 1 bit from $\overline{QX}$ to $\overline{QXP}$.
Simultaneously, half a bit flows from $\overline{X}$ to $\overline{P}$, as $f(t) + f^\dagger(t)$ becomes almost entirely supported within the $P$ region.

\begin{figure}[tbp]
	\centering
	\includegraphics[width=\columnwidth]{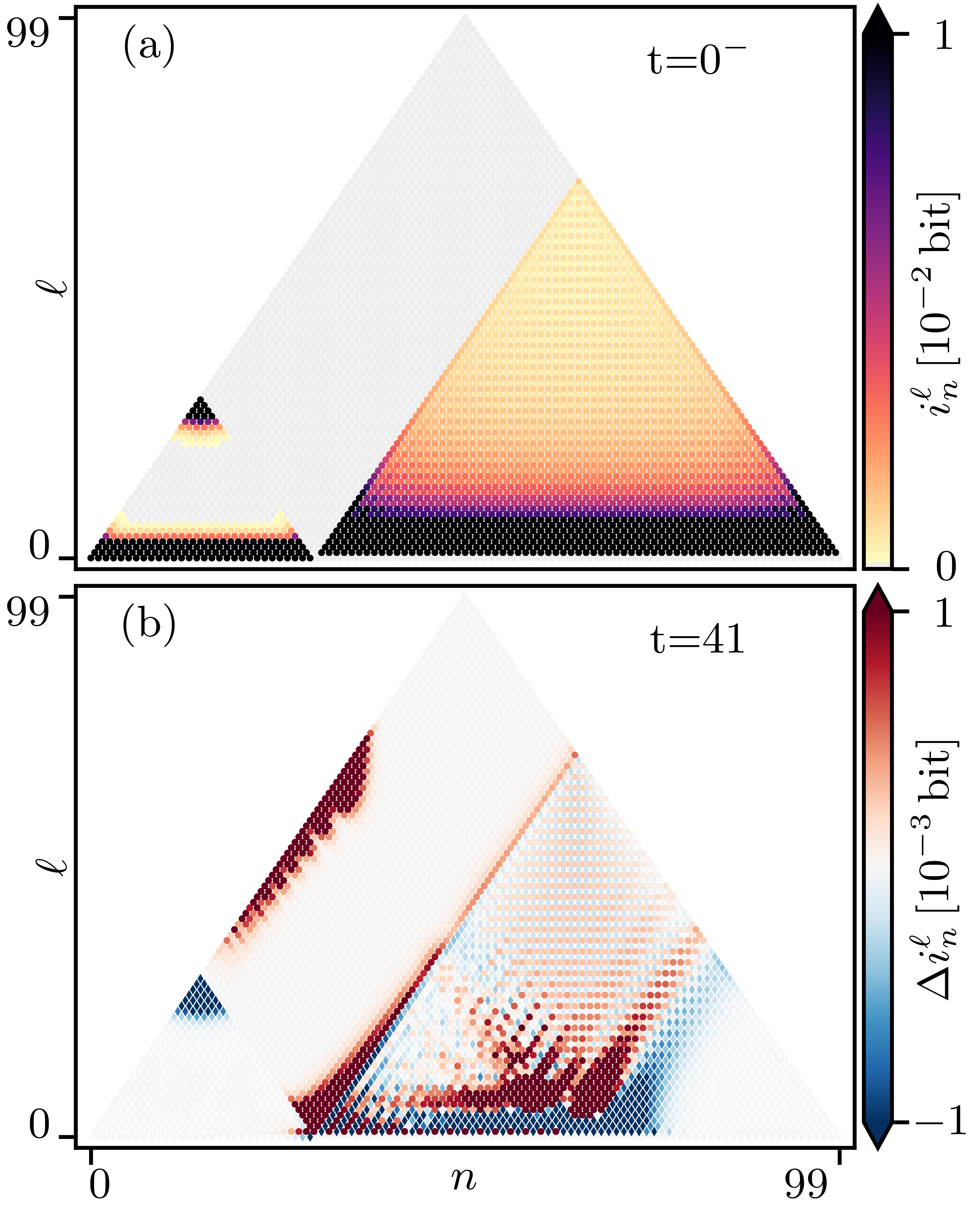}
	\caption{(a)-(b) Time evolution of local information after quenching the tunnel coupling between a topological Kitaev chain with finite chemical potential Eq.~\eqref{eq:HamKitaevFiniteMu} and a critical tight-binding chain.
    In (b) red dots (blue diamonds) indicate an increase (decrease) of local information.
    The parameters of the Kitaev chain are chosen away from the topological sweet spot: $\mu/\tau=0.6$ and $\tau_p/\tau=0.01$ ($l_Q=30,l_X+l_P=70$).} 
	\label{fig:IL_MZM_Deviation_TSS}
\end{figure}

The effective Hamiltonian description holds under three main conditions.
First, $l_Q \gg \lambda$, where $\lambda$ is the pre-quench correlation decay length in $Q$ defined in Ref.~\cite{artiaco2024universal}.
Second, the $Q$ region is initially in the topological phase.
Third, the energy scale $\tau$ of the $Q$ region is much larger than that of the rest of the system.
These conditions apply not only to the sweet-spot Hamiltonian~\eqref{eq:HamKitaevChainAtTSS}, but also to a broad class of models---including those with additional superconducting, chemical potential, or interaction terms of order $\tau$~\cite{Fendley2016}.
In such cases, $\lambda$ increases, and the localized bit at the top of $\overline{Q}$ in Fig.~\ref{fig:il_MZM_teleportation_SummaryPlots} broadens over a region of width $\approx \lambda$.
This is illustrated in Fig.~\ref{fig:IL_MZM_Deviation_TSS}, where we add a chemical potential to the sweet-spot Hamiltonian:
\begin{align}
H_{\mathrm{KC}}^Q= \sum_{i=0}^{l_Q-1} \mu \, c_i^\dagger c^{\phantom{\dagger}}_i + \frac{\tau}{2}\sum_{i=0}^{l_Q-2} \left( c_i^\dagger c^{\phantom{\dagger}}_{i+1} + c^{\phantom{\dagger}}_i c^{\phantom{\dagger}}_{i+1}+\text{H.c.} \right),
\label{eq:HamKitaevFiniteMu}
\end{align}
with $\mu=0.6\tau$.

The correction terms in the effective Hamiltonian~\eqref{eq:effHamQuenchProbeTB} of order $\mathcal{O}(\tau_{t/p}^2/\tau)$ become relevant when the energy scale $\tau$ of the $Q$ region is no longer much larger than that of the rest of the system.
In such cases, the effective Hamiltonian may no longer provide an accurate description, and the main features of the local information flow discussed above can qualitatively change.
Specifically, (i) the 1 bit associated with topological edge correlation can mix with long range packets emitted from the quenched interface and (ii) local information at small scale can flow from the bulk into the critical tight-binding probe and vice versa.
Thus, at long times one cannot distinguish between the bulk and the MZM contribution to the local information flow.

However, at the sweet spot, that is, as long as the Kitaev chain is governed by the Hamiltonian~\eqref{eq:HamKitaevChainAtTSS}, the validity of the effective description~\eqref{eq:effHamQuenchProbeTB} does not depend on the condition $\tau \gg \tau_{t/p}$.
The only difference in that regime is that the site at $l_Q - 2$ becomes correlated with the $XP$ region.
This is because the sweet-spot Hamiltonian possesses a set of local integrals of motion, $\{d_i^\dagger d_i\}_{i=0}^{l_Q - 2}$, all of which remain conserved after the quench except for the one near the interface ($i = l_Q - 2$).
Even for the coupling parameters used in Figs.~\ref{fig:il_MZM_teleportation_SummaryPlots}(b)-(c) we observe a very small but finite amount of local information that is located at the interface $(i^2_9\approx0.0014\text{ bit})$ due to the coupling of the tight-binding chain to the site $i=l_Q-2$.
Similarly, moving away from the sweet spot but introducing disorder into the Kitaev chain restores the local integrals of motion~\cite{parameswaran2018many,laflorencie2022topological}, and in the large $l_Q$ limit the resulting dynamics again resembles the sweet-spot scenario.

\section{Conclusions}
\label{sec:conclusions}

In this article, we demonstrated the use of the information lattice to fully characterize quantum quenches.
The information lattice decomposes the total information in a system into local components, termed local information, which quantifies how information is distributed across different scales and spatial locations.
Thanks to the addition of an extra dimension encoding the scale, within the information lattice, information behaves like a hydrodynamic quantity: while unitary time evolution conserves the total information in a system, local information flows between information lattice sites through well-defined local information currents.

As an example, we examined three local quench protocols in noninteracting fermionic chains.
In the first quench, we investigated the propagation of a fermionic particle released at the central site of an empty tight-binding chain.
We followed its time evolution through the flow of local information and analyzed the development of both short- and long-range correlations.
Correlations in the system developed over time due to the spreading of the initially localized wave function of the particle at the central site.
The information lattice fully characterized the spatial and scale extend and spreading of this wave function through local information.
As the wave function spread equally to the left and right, the two halves of the physical chain became maximally entangled. 
In the second quench protocol, we considered a critical tight-binding chain with a potential barrier at the central site.
We prepared the system in the ground state in the presence of the barrier and induced out-of-equilibrium dynamics via the instantaneous removal of the barrier at $t=0$.
We characterized the specific features of quench protocols within critical chains, which are due to the presence of local information in the initial state not only at short but also at larger scales.
This gives rise to information interface effects visible via the information lattice.

Finally, we analyzed a quench protocol in which a topological Kitaev chain in its ground state is suddenly coupled to a critical tight-binding chain in its ground state.
The local decomposition of the total information obtained through the information lattice allowed us to identify the specific features of the local information flow associated with Majorana edge modes.
Fractional signatures in the von Neumann entropy due to the presence of localized Majorana edge mode have been previously reported~\cite{sela2019detecting,bauer2023quench}.
However, exploiting the information lattice, we distinguished two distinct origins thereof.
One arises from an interface effect between the topological and critical chain, localizing a fractional amount of information.
The other stems purely from a non-equilibrium process, driven by a propagating Majorana mode induced by the quench.
Using the microscopic flow of information provided by the information lattice, we analytically derived the origin and properties of both mechanisms.

While we focused in this work on local quenches in noninteracting fermionic systems, our framework is entirely general and applies to global quenches, interacting systems, and systems subject to dephasing and dissipation.
In interacting systems, information tends to propagate from short to large scales, as dictated by the second law of thermodynamics.
This principle has been exploited in Refs.~\cite{klein2022time,artiaco2024efficient,harkins2025nanoscale} to develop efficient time-evolution schemes for many-body dynamics.
For generic interacting systems, this suggests a breakdown of the clear separation between short- and long-range correlations observed in the quench protocols analyzed in this work.
The information lattice provides a powerful and unified perspective on out-of-equilibrium quantum dynamics.
By acting as an “information microscope” and enabling the tracking of correlations at a local level, the information lattice could play a crucial role in advancing our understanding of out-of-equilibrium phenomena---for instance, in contexts presenting the interplay between unitary time evolution and measurements in measurement-induced phase transitions~\cite{li2018quantum,skinner2019measurement}.
\section*{Data availability}
The data that support the findings of this article are openly available \cite{bauer2025rawdata}.\\

\section*{Acknowledgements}
This work was supported by the DFG-SFB 1170 (Project-ID: 258499086) and EXC2147 ct.qmat (Project-ID: 390858490), the
European Research Council (ERC) under the European Union’s Horizon 2020 research and innovation program
(Grant Agreement No.~101001902) and the Knut and Alice Wallenberg Foundation (KAW) via the project Dynamic Quantum Matter (2019.0068).
T. K. K. acknowledges funding from the Wenner-Gren Foundations.
The computations were enabled by resources provided by the National Academic Infrastructure for Supercomputing in Sweden (NAISS), partially funded by the Swedish Research Council through grant agreement no. 2022-06725.

\bibliography{Ref.bib}

\end{document}